\documentclass[11pt,a4paper]{article}
\pdfoutput=1
\usepackage[utf8]{inputenc}
\usepackage{jheppub}            
\makeatletter
\gdef\@fpheader{}               
\makeatother
\usepackage{solxpaper}          

\begin{document}

\title{\solx: supergravity theories/solutions as code\\[6pt]{\large\itshape with an agentic workflow for machine-verifiable physics}}

\author[a]{Vasil Dimitrov}
\affiliation[a]{INRNE, Bulgarian Academy of Sciences, Tsarigradsko Chaussee 72, 1784 Sofia, Bulgaria}
\emailAdd{vasildimi@gmail.com}

\abstract{%
A supergravity solution in the literature is an artifact that cannot be verified without re-deriving it: conventions might not be clearly stated, papers have typos, algebra is tedious. \solx is a \mma package that turns an \xact environment, whatever physics it may describe, into a stored, machine-verifiable record: code that replays in a fresh kernel, carries its own verification, and can later be fetched for re-use or extension. The shipped examples are supergravity theories and their solutions, but the scope is anything \xact reaches: abstract tensors, differential forms, spinors in any dimension. We follow one published solution through the entire workflow, benchmark the package's component engine against \xact's, and report on three blind runs of an agentic workflow in which an AI coding agent, given the package and a small toolkit, correctly codified a recent paper in under three hours. A database of such records is useful to researchers today and gives agents worked examples to learn from. The package acts as an internal verifier during the thought process of an agent: a claim is checked in the kernel before it is made.
\par\vspace{2ex}\noindent\textsc{Program summary:}\par\nopagebreak\vspace{1ex}
\begin{tabular}[t]{ll}
Program title: & \solx \\
Version: & 2.1.0 \\
Programming language: & \href{https://www.wolfram.com/mathematica/}{\mma} (version 13.3 or later) \\
Dependencies: & \href{http://www.xact.es}{\xact} \\
Obtainable from: & \href{https://github.com/waskou/SolutionsX}{\nolinkurl{github.com/waskou/SolutionsX}} \\
 & or the \href{https://resources.wolframcloud.com/PacletRepository/resources/VasilDimitrov/SolutionsX/}{Wolfram Paclet Repository} \\
License: & \href{https://www.gnu.org/licenses/agpl-3.0.html}{AGPL-3.0-or-later} (code) \\
 & \href{https://creativecommons.org/licenses/by/4.0/}{CC BY 4.0} (bundled data) \\
Mascot: & \raisebox{\dimexpr\ht\strutbox-\height\relax}{\includegraphics[width=5cm]{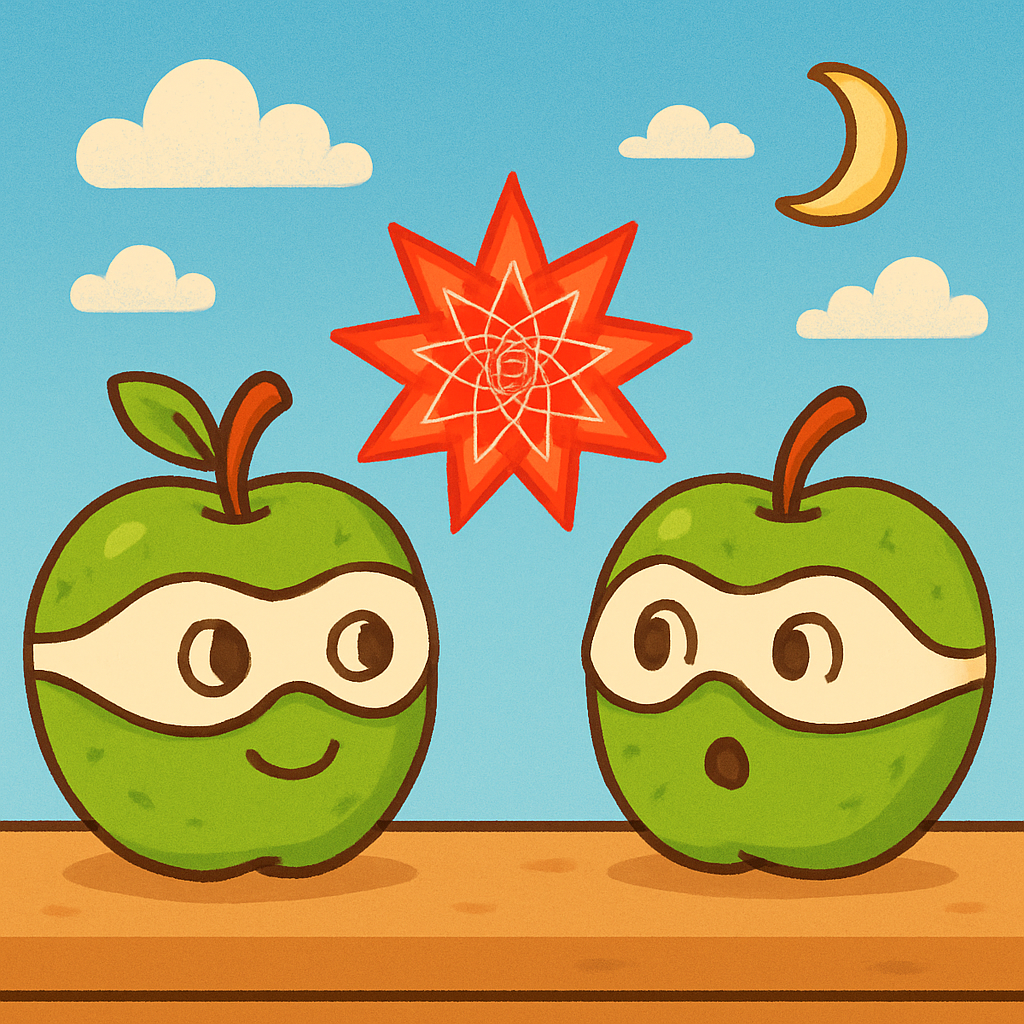}}
\end{tabular}}

\maketitle

\section{Introduction}
\label{sec:intro}

There is a plethora of computer algebra systems (CAS) that can handle differential geometry, tensor calculus and explicit-coordinates-calculations. A person working with supersymmetry, or more generally one who wants to couple fermions to gravity --- dynamical or as a fixed background --- requires a CAS that can also handle spinors, gamma matrices, generalized connections, \dots, preferably in arbitrary dimensions. We are aware of the following pieces of software that
achieve that in various capacity:
\begin{itemize}
  \item \href{http://www.xact.es}{\xact}~\cite{Martin-Garcia:2008ysv}: a free suite of packages
    for abstract tensor calculus (\xtensor), component calculus (\xcoba),
    exterior calculus (\xterior), spinors and fermions
    (\fieldsx~\cite{Frob:2020gdh}), and more,
    running on top of the commercial \mma.
  \item \href{https://cadabra.science}{\texttt{Cadabra2}}~\cite{Peeters:2018dyg}: a free standalone
    CAS designed specifically for field-theory problems.
  \item \href{https://sagemanifolds.obspm.fr}{\texttt{SageManifolds}}~\cite{Gourgoulhon:2014ywa}:
    differential geometry and tensor calculus within the free, Python-based
    \texttt{SageMath} system.
  \item
    \href{https://www.maplesoft.com/support/help/maple/view.aspx?path=DifferentialGeometry}{\texttt{DifferentialGeometry}}~\cite{Anderson:2011mb}:
    tensor calculus and Newman--Penrose/spinor formalisms within the
    commercial \texttt{Maple}.
\end{itemize}
\solx builds upon \xact. It is a combination of three key functionalities
\begin{enumerate}[(i)]
  \item Ability to work with spinors and fermionic fields in arbitrary dimension --- already present in \xact, most importantly through \fieldsx
  \item An open database of theories and their solutions, whose entries are \textit{verifiable}, \textit{re-usable} and subject to \textit{human/agent contributions}
  \item \textit{Fast} and \textit{flexible} computations of tensor components --- for any type of tensor, carrying indices in any vector bundle --- with ability to parallelize\footnote{The words ``fast'' and ``flexible'' will be quantified in \cref{sec:bench}}
\end{enumerate}
Essentially, \solx is a ``wrapper'' over an existing \xact environment\footnote{By that we mean all the symbols that the user defines using \xact, together with any computations done in a given kernel session}, such that the environment can be stored and re-played at will in a new kernel. As such it automatically ports the capabilities of \xact and its associated sub-packages --- and it is deliberately agnostic about what the environment describes. In practice a stored record is one of two kinds. A \textit{theory} is an abstract environment: a manifold, a metric with no chart attached, the field content, and the distinguished expressions that define the physics --- equations of motion, supersymmetry variations, Killing-spinor conditions, the latter built on \fieldsx. A \textit{solution} is the same environment specialized to explicit coordinates --- charts, component metrics, fluxes, Killing spinors --- together with the replayed computations that verify it. Neither notion is tied to supergravity: a QFT with fermions, supersymmetric or not, on a fixed curved background is the same kind of environment with the metric simply non-dynamical, and is covered automatically. 
The examples throughout the paper are drawn from supergravity and holography, since that is the expertise of the author, not because the system cannot handle more general setups.
Of course \solx, like \xact, lives on the geometric side of such a problem --- backgrounds, spinors, variations --- and does not attempt loop calculations. In all cases the user computes once and stores for the future.

Before the broader vision, the pragmatic question: what does a user gain the day they install \solx? Fetching a stored theory replaces re-deriving it: conventions, field content and defining equations arrive as code that loads in seconds. A finished calculation becomes an entry that replays on a fresh kernel, so it can be re-checked, extended, or handed over years later. Stored entries work together: in \cref{sec:life} one black hole is obtained as a limit of another, and a holographic computation is spread across five cooperating entries. Finally, the package is what turns an AI agent into a useful collaborator on such problems, which is the subject of \cref{sec:agents}.

Now for the broader vision, with more users and more worked out examples the aim is to have:
\begin{quote}\itshape
A database of theories and solutions which are verifiable code, not merely arXiv prose
\end{quote}
The idea behind this is clear and mathematics is already demonstrating the utility. Once statements live in a form a machine can check,
correctness stops being a matter of trust (or time spent verifying) and collaboration scales up:
\href{https://lean-lang.org}{\texttt{Lean}}'s library
\href{https://leanprover-community.github.io}{\texttt{mathlib}} now counts
over $280\,000$ verified theorems from nearly $800$ contributors, and the
\href{https://leanprover-community.github.io/blog/posts/lte-final/}{Liquid
Tensor Experiment} formalized a central theorem of condensed mathematics
within eighteen months of Scholze's challenge. The same substrate admits
machines as contributors:
\texttt{AlphaProof}~\cite{Hubert:2025AlphaProof},
searching over \texttt{Lean} proofs, reached silver-medal standard at the
2024 International Mathematical Olympiad --- possible precisely because
every candidate proof could be verified. In 2026 the pace of progress is on a different level: working through
\href{https://www.erdosproblems.com}{a curated database} of nearly a
thousand Erd\H{o}s problems, agents delivered the first fully autonomous
resolution of one~\cite{Sothanaphan:2026Erdos728} (\#728, January 2026,
proof in \texttt{Lean}), and by
\href{https://www.quantamagazine.org/why-the-legendary-erdos-problems-are-falling-to-ai-20260803/}{mid-2026}
the community tracker recorded AI contributions to over a hundred more ---
genuinely new proofs, including a disproof of the unit-distance
conjecture posed in 1946~\cite{Alon:2026UnitDistance}, next to many
rediscoveries of results buried in the literature.
We aim at a hep-th analogue currently restricted to
supergravity and QFT on a fixed background: a stored theory or solution is code that re-runs,
so a contribution --- from a colleague or from an agent --- is machine verifiable. This removes a familiar bottleneck of
chasing conventions/typos in papers and allows one to build upon a trusted source. The long-term goal is that a reliable database of theories and solutions will not only be useful to human researchers, but will allow agents to ``learn from the existing code'' and contribute meaningfully to novel research. The division of labour deserves to be stated plainly: there is no AI inside the package. \solx is deterministic computer algebra; the AI enters as a user of it, reading the corpus for worked examples, authoring new entries, and consulting the kernel as an
independent verifier.

In addition to \xact's built in functionality, \solx defines a handful of functions that the author deemed important in their own workflow which extend \xact's capabilities. These additions are mainly concerned with converting between differential forms representation and tensor representation of geometric objects, see \sx{MakeArray}, \sx{MakeMetric}, \sx{MakeForm}, \sx{ExpandForms}, \sx{WedgeCoeff}, \sx{ComputeDiffs}.

The paper is organized as follows. \Cref{sec:installation} contains the installation instructions. \Cref{sec:life} follows a single record through its life: how a theory and a solution are entered, computed, verified, stored and reused. \Cref{sec:bench} benchmarks the component engine against \xcoba's native one on two five-dimensional black-hole records. \Cref{sec:agents} presents the agent toolkit and three blind runs in which an agent re-works three published physics problems. \Cref{sec:conclusions} classifies types of problems that the author deemed amenable to codification. \Cref{app:mma} collects the \mma notation the paper's cells rely on, \cref{app:surface} lists every public symbol with its usage message, and \cref{app:kit} documents the configuration of the agent toolkit.

\section{Installation}
\label{sec:installation}
Installation is in three steps: the prerequisites, the package itself, and a first-session wizard. Steps 1 and 3 are identical for everyone. Step 2 comes in two routes, and the choice depends on whether you want the package alone or the whole repository around it.

\paragraph{Step 1: prerequisites (both routes)} \solx runs in \mma, version 13.3 or newer, and builds on two packages that have to be installed first. Both are installed by placing files in a specific folder, and \mma itself tells you where that folder is: evaluate in a notebook
\begin{wlin}
FileNameJoin[{\$UserBaseDirectory, "Applications"}]
\end{wlin}
\noindent and the printed path is the folder\footnote{Typically \code{\textasciitilde/Library/Mathematica/Applications} on macOS and \code{\textasciitilde/.Mathematica/Applications} on Linux; newer installations say \code{Wolfram} in place of \code{Mathematica}.}, called \code{Applications} below. Three items go into it:
\begin{itemize}\itemsep2pt
  \item \href{http://www.xact.es}{\xact}. Download the bundle from its \href{http://www.xact.es/download.html}{download page} and unzip it; the result is a folder named \code{xAct}. Move that folder into \code{Applications}.
  \item \href{https://github.com/mfroeb/FieldsX}{\fieldsx}~\cite{Frob:2020gdh}. On its repository page press the \code{Code} button and choose \code{Download ZIP}; unzip the download. From the resulting folder copy the single file \code{FieldsX.m} into the \code{xAct} folder of the previous item, so that now it is located at \code{Applications/xAct/FieldsX.m}.
  \item \code{Multisets.m}. \fieldsx relies on this one further file, downloaded from the \href{https://library.wolfram.com/infocenter/MathSource/8115/}{Wolfram Library Archive} and placed directly in \code{Applications}, beside the \code{xAct} folder, at \code{Applications/Multisets.m}.
\end{itemize}
\important{} \code{FieldsX.m} should sit inside the \code{xAct} folder, which is inside the \code{Applications} folder; \code{Multisets.m} should sit beside the \code{xAct} folder, directly in the \code{Applications} folder.

\paragraph{Step 2, route A: the paclet} The package, its documentation and the curated example entries ship together, and installing them is one line, evaluated in any notebook
\begin{wlin}
PacletInstall["VasilDimitrov/SolutionsX"]
\end{wlin}
\noindent which fetches the package from the \href{https://resources.wolframcloud.com/PacletRepository/resources/VasilDimitrov/SolutionsX/}{Wolfram Paclet Repository}.

\noindent \important{} The install line above needs \href{https://www.wolframcloud.com/}{\nolinkurl{wolframcloud.com}} to be reachable. At the time of writing the Wolfram Cloud is undergoing an internal update. Until that update is complete, use route B.

\paragraph{Step 2, route B: the repository} Take this route if you want more than the package: the data directory with the published corpus (including the agent-authored entries of \cref{sec:agents}), the agent toolkit \code{agent/}, and the package source. It is also the starting point for contributing entries of your own. The repository lives at \href{https://github.com/waskou/SolutionsX}{\nolinkurl{github.com/waskou/SolutionsX}}, and installing it means getting a copy of that folder onto your computer, then running its setup script. Getting the copy, with or without git:
\begin{itemize}\itemsep2pt
  \item With git (recommended). Open a terminal and run
  \shellline{git clone https://github.com/waskou/SolutionsX.git}
  The folder \code{SolutionsX} appears; a later \code{git pull} inside it fetches updates, and contributing back runs through git as well.
  \item Without git. Open the page in a browser, press the \code{Code} button, and choose \code{Download ZIP}. Unzip the downloaded file; the result is the same folder, named \code{SolutionsX-main}\footnote{The \code{-main} suffix is GitHub's naming and is harmless; if you prefer to rename or move the folder, do it before running the setup script, which records the folder's location.}. Move it wherever you want to keep it. Beware that the ZIP is a plain snapshot with no update mechanism: to update you download a fresh copy and move your own \code{Data} alias folder into it by hand, which is why we recommend the git route.
\end{itemize}
Now run the setup script from inside that folder: open a terminal, type \code{cd} followed by a space and the folder's path, press enter, and then run
\shellline{wolframscript -file setup.wls}
\noindent The \code{wolframscript} executable ships with \mma. The script builds the paclet from the repository, installs it, and points your \solx configuration at the repository's \code{Data} directory.

\noindent \important{} The repository folder itself stays where you put it: unlike the prerequisites of step 1, nothing is copied or moved into \code{Applications}.

\paragraph{Step 3: the first session (both routes)} Load the package with
\begin{wlin}
Needs["VasilDimitrov`SolutionsX`"]
\end{wlin}
\noindent and run the welcome wizard once
\begin{wlin}
Welcome[]
\end{wlin}
\noindent The wizard sets up your local storage: where the data lives (on route B this is already the repository's \code{Data} directory), the alias under which your own entries will be saved, and which curated entries to copy under it. The answers are remembered, so every later session starts directly with \wlref{Needs}\code{["VasilDimitrov\textasciigrave{}SolutionsX\textasciigrave{}"]}. With the package loaded, one of the shipped solutions is fetched and loaded with
\begin{wlin}
(!\free{sol}!) = GetData["Curated/Sol__4dL__Black-hole__Global"];
Load@(!\free{sol}!)
\end{wlin}
\noindent and if the wizard copied the curated entries under your own alias, the bare name suffices
\begin{wlin}
(!\free{sol}!) = GetData["Sol__4dL__Black-hole__Global"];
\end{wlin}
\noindent Authoring a new solution starts with
\begin{wlin}
NewData[]
\end{wlin}
\noindent which opens a new notebook with the initialization cell already in place. The available entries are browsed interactively with
\begin{wlin}
ShowData[]
\end{wlin}
\noindent and the in-package documentation opens with
\begin{wlin}
SolutionsXHelp[]
\end{wlin}
\noindent The usage message of a symbol prints directly with
\begin{wlin}
?Compute
\end{wlin}
\noindent and the \textcircled{\scriptsize i} icon in the message opens the symbol's documentation page, exactly as for the built-in \mma symbols.

\section{The life of a solution}
\label{sec:life}
In this section we exemplify how a solution gets created, what information about it is stored and where, how computations are done, and how the solution relates to other theories and solutions. The presentation favours narrative structure, so not everything needed to reproduce the example in a \mma kernel is explicitly shown. \Cref{app:mma} collects the \mma idioms the cells below rely on. The section mainly follows the example entries:
\begin{itemize}
  \item \sxdata{Curated/Thr__5dL__Minimal-gauged}
  \item \sxdata{Curated/Sol__5dL__CCLP-susy__Orthotoric}
\end{itemize}

\subsection{Fetching the parent theory}
What counts as a \textit{theory} (\texttt{Thr}) and what counts as a \textit{solution} (\texttt{Sol}) is determined by the rudimentary question: Does the \xact environment define a local chart? If no, it is a theory; if yes, a solution. Otherwise the distinction is merely semantics.

We begin by fetching the already stored theory. In this case this is 5d minimal supergravity whose Lorentzian bosonic action is
\begin{align}\label{eq:action}
  S ={}& \frac{1}{16\pi G_5} \int \qty[\qty(R - \frac{12}{\ell_5^2}) \star 1 - \frac{C_x^2}{2} F \wedge \star F - \frac{C_x^3}{3 \sqrt{3}} A \wedge F \wedge F] \,,
\end{align}
where $G_5$ is Newton's constant, $\ell_5$ is the $\text{AdS}_5$ scale and $C_x$ is the placeholder for a possibly different normalization of the gauge field. We fetch and load it with
\begin{wlin}
(!\free{sol}!) = GetData["Curated/Thr__5dL__Minimal-gauged"];
Load@(!\free{sol}!)
\end{wlin}
\begin{wlpanel}
(!\wlplain{Loaded:}!)
@[ScriptCapitalM], @[DoubleStruckCapitalT]@[ScriptCapitalM], @[ScriptM], @[ScriptN], @[ScriptO], @[ScriptP], @[ScriptQ], @[ScriptR], @[ScriptS], @[ScriptT], @[ScriptU], @[ScriptV], @[ScriptW], @[ScriptX], @[ScriptY], @[ScriptZ], PerturbationParametergg, gg, cd, epsilongg, Torsioncd, Christoffelcd, Riemanncd, Riccicd, RicciScalarcd, Einsteincd, Weylcd, TFRiccicd, Kretschmanncd, SymRiemanncd, Schoutencd, SchoutenCCcd, EinsteinCCcd, Detgg, Perturbationgg, Koszulcd, CovDcd, @[Eta]@[Eta], @[DoubleStruckCapitalF]@[ScriptCapitalM], ee, Detee, epsilon@[Eta]@[Eta], @[GothicA], @[GothicB], @[GothicC], @[GothicD], @[GothicE], @[GothicF], @[GothicG], @[GothicH], @[GothicI], @[GothicJ], @[GothicK], @[GothicL], @[DoubleStruckCapitalS]@[ScriptCapitalM], AChristoffelcd, FRiemanncd, Gammagg1, Gamma@[Eta]@[Eta]1, Gammagg0, Gamma@[Eta]@[Eta]0, Gammagg2, Gamma@[Eta]@[Eta]2, Gammagg3, Gamma@[Eta]@[Eta]3, Gammagg4, Gamma@[Eta]@[Eta]4, Gammagg5, Gamma@[Eta]@[Eta]5, @[DoubleStruckCapitalA], @[DoubleStruckCapitalB], @[DoubleStruckCapitalC], @[DoubleStruckCapitalD], @[DoubleStruckCapitalE], @[DoubleStruckCapitalF], @[DoubleStruckCapitalG], @[DoubleStruckCapitalH], @[DoubleStruckCapitalI], @[DoubleStruckCapitalJ], @[DoubleStruckCapitalK], @[DoubleStruckCapitalL], @[Omega]@[Omega], Riemann@[Omega]@[Omega], Ricci@[Omega]@[Omega], RicciScalar@[Omega]@[Omega], PDflat, flat, TorsionPDflat, ChristoffelPDflat, RiemannPDflat, RicciPDflat, AChristoffelPDflat, FRiemannPDflat, etaUpflat, etaDownflat, flat0, flat1, flat2, flat3, flat4, KoszulPDflat, CovDPDflat, PDspin, spin, TorsionPDspin, ChristoffelPDspin, RiemannPDspin, RicciPDspin, AChristoffelPDspin, FRiemannPDspin, etaUpspin, etaDownspin, spin1, spin2, spin3, spin4, KoszulPDspin, CovDPDspin, TA, A, TF, F, @[Epsilon]@[Epsilon], bar@[Epsilon]@[Epsilon], G5, l5, Cx
(!\wlplain{Info[...]}!)
\end{wlpanel}
\noindent The output is a live panel: clicking on a symbol shows its \wlref{Information} panel, and clicking again closes it. Only a small part of what is listed was fed in by hand when \sxdata{Curated/Thr__5dL__Minimal-gauged} was authored:
\begin{itemize}\itemsep2pt
  \item \code{\$manifold} --- the spacetime $\ensuremath{\mathcal{M}}$: five-dimensional, with abstract indices $\curly{m},\curly{n},\dots,\curly{z}$ on its (automatically created) tangent bundle \code{Tangent}$\ensuremath{\mathcal{M}}$, printed $\mathbb{T}\ensuremath{\mathcal{M}}$, and the \code{Name} \wlstr{Minimal-gauged}
  \item \code{\$metric} --- the metric \code{gg}, declared with \wlopt{Signature}{-1} and no chart, and its Levi-Civita connection \code{cd}
  \item \code{\$frame} --- the vielbein \wlcall{ee}{-\curly{m},\mathfrak{a}} and the flat metric \wlcall{\ensuremath{\eta\eta}}{-\mathfrak{a},-\mathfrak{b}}, where $\mathfrak{a},\mathfrak{b},\dots,\mathfrak{l}$ are the indices on the automatically created vector bundle \code{Frame}$\ensuremath{\mathcal{M}}$, printed $\mathbb{F}\ensuremath{\mathcal{M}}$.
  \item \code{\$spinStructure} --- over \code{gg} with spinor indices $\mathbb{A},\mathbb{B},\dots,\mathbb{L}$ for the vector bundle \code{Spin}$\ensuremath{\mathcal{M}}$, printed $\mathbb{S}\ensuremath{\mathcal{M}}$\footnote{Now a curved $\gamma$-matrix is called as, eg. \wlcall{Gammagg2}{-\curly{m},-\curly{n},-\mathbb{A},\mathbb{B}}  and a flat $\gamma$-matrix is called as \wlcall{Gamma\ensuremath{\eta\eta}2}{-\mathfrak{a},-\mathfrak{b},-\mathbb{A},\mathbb{B}}; the trailing digit is the rank, and the heads run \code{Gammagg0} to \code{Gammagg5} in five dimensions}
  \item \code{\$spinConnection} --- $\omega\omega[-\curly{m},-\mathfrak{a},-\mathfrak{b}]$ associated to the Levi-Civita connection \code{cd}.
  \item \code{\$basis} --- the two bases: \code{flat} on the frame bundle $\mathbb{F}\ensuremath{\mathcal{M}}$, colored \basisswatch{sxflat}, numbered $0\dots4$; and \code{spin} on the spin bundle $\mathbb{S}\ensuremath{\mathcal{M}}$, colored \basisswatch{sxspin}, numbered $1\dots4$\footnote{\sx{ToBases} now converts \wlcall{Gamma\ensuremath{\eta\eta}2}{-\mathfrak{a},-\mathfrak{b},-\mathbb{A},\mathbb{B}} into \wlcall{Gamma\ensuremath{\eta\eta}2}{\qty{-\mathfrak{a},-\code{flat}},\allowbreak\qty{-\mathfrak{b},-\code{flat}},\allowbreak\qty{-\mathbb{A},-\code{spin}},\allowbreak\qty{\mathbb{B},\code{spin}}} and it prints as {\normalsize$\gamma[\eta\eta]\indices{_{\iflat{\mathfrak{a}}}_{\iflat{\mathfrak{b}}}_{\ispin{\mathbb{A}}}^{\ispin{\mathbb{B}}}}$}, each index wearing its basis colour}
    \item \code{\$form} --- the gauge potential \code{A[]} (differential form of degree one) and its field strength \code{F[]} (differential form of degree two). Each one arrives with a tensor partner, \wlcall{TA}{-\curly{m}} and \wlcall{TF}{-\curly{m},-\curly{n}}, automatically created by \solx\footnote{Notice the \code{[]} after the form symbols; this is \xterior's way to denote forms that carry no indices. For an example with tensor-valued forms consult \sxdata{Curated/Thr__5dL__STU-gauged}.}
  \item \code{\$spinor} --- the Dirac spinor $\epsilon\epsilon[-\mathbb{A}]$, and with it its bar.
  \item \code{\$constant} --- $G_5$, $\ell_5$ and $C_x$, the constants of the action above.
\end{itemize}
The rest is everything \xact automatically defines. The entry's own name is ``deduced'': no chart makes it a \code{Thr}, \wlopt{Dimension}{5} with \wlopt{Signature}{-1} makes it \code{5dL}, and \code{sol[\$manifold, }$\ensuremath{\mathcal{M}}$\code{, Name]} is appended to that. Aside from that \sxdata{Curated/Thr__5dL__Minimal-gauged} calls \code{IncludeTo[sol[\$set]]} to store the equations of motion
\begin{wlin}
(!\free{\ensuremath{\ensuremath{\mathcal{E}}}gg}!)[-(!\fre{\curly{m}}!), -(!\fre{\curly{n}}!)]
\end{wlin}
\begin{wloutm}
\[
  \frac{4\,\code{gg}_{\curly{m}\curly{n}}}{\code{l5}^2}
  + \code{R}[\nabla]_{\curly{m}\curly{n}}
  - \frac{\code{Cx}^2}{2}\,
    \code{TF}\indices{_{\curly{m}}^{\curly{o}}}\,\code{TF}_{\curly{n}\curly{o}}
  + \frac{\code{Cx}^2}{12}\,
    \code{gg}_{\curly{m}\curly{n}}\,
    \code{TF}_{\curly{o}\curly{p}}\,\code{TF}^{\curly{o}\curly{p}}
\]
\end{wloutm}
\begin{wlin}
(!\free{\ensuremath{\ensuremath{\mathcal{E}}}TA}!)[-(!\fre{\curly{m}}!)]
\end{wlin}
\begin{wloutm}
\[
  -\frac{\code{Cx}\,
    \epsilon\code{gg}_{\curly{m}\curly{n}\curly{o}\curly{p}\curly{q}}\,
    \code{TF}^{\curly{n}\curly{o}}\,\code{TF}^{\curly{p}\curly{q}}}{4\sqrt{3}}
  + \nabla_{\curly{n}}\code{TF}\indices{_{\curly{m}}^{\curly{n}}}
\]
\end{wloutm}
\noindent which were previously derived from the Lagrangian using \xact's built-in functions; and the Killing spinor equation
\begin{wlin}
(!\fre{\delta\psi\psi}!)[-(!\fre{\curly{m}}!), -(!\fre{\mathbb{A}}!)]
\end{wlin}
\begin{wloutm}
\begin{align*}
  &-\frac{\iu\sqrt{3}\,\code{Cx}\,\code{TA}_{\curly{m}}\,
      \epsilon\epsilon_{\mathbb{A}}}{2\,\code{l5}}
   -\frac{\gamma[\code{gg}]\indices{_{\curly{m}}_{\mathbb{A}}^{\mathbb{B}}}\,
      \epsilon\epsilon_{\mathbb{B}}}{2\,\code{l5}}
   +\frac{\iu\,\code{Cx}\,
      \gamma[\code{gg}]\indices{^{\curly{n}}_{\mathbb{A}}^{\mathbb{B}}}\,
      \code{TF}_{\curly{m}\curly{n}}\,
      \epsilon\epsilon_{\mathbb{B}}}{2\sqrt{3}} \\[2pt]
  &\quad-\frac{\iu\,\code{Cx}\,
      \gamma[\code{gg}]\indices{_{\curly{m}}_{\curly{o}}_{\curly{n}}_{\mathbb{A}}^{\mathbb{B}}}\,
      \code{TF}^{\curly{o}\curly{n}}\,
      \epsilon\epsilon_{\mathbb{B}}}{8\sqrt{3}}
   +\frac{1}{4}\,
      \gamma[\eta\eta]\indices{^{\mathfrak{b}}^{\mathfrak{a}}_{\mathbb{A}}^{\mathbb{B}}}\,
      \epsilon\epsilon_{\mathbb{B}}\,
      \omega\omega_{\curly{m}\mathfrak{b}\mathfrak{a}}
   +\partial_{\curly{m}}\epsilon\epsilon_{\mathbb{A}}
\end{align*}
\end{wloutm}
\noindent which one can in principle derive from scratch using \fieldsx, once the fermionic terms are supplied to \eqref{eq:action}. What is useful about this setup is that \sx{Load} loaded that entire \xact environment. The theory and its defining expressions are derived once, then called upon whenever needed. For a more involved pre-built theory, one carrying vector multiplets, see \sxdata{Curated/Thr__5dL__STU-gauged}.

\subsection{Inputting the solution}
\label{sec:input}
Having loaded the theory, we ``append'' to it the data that specifies a concrete solution. In the example below we build up towards a known solution; one can also build up an ansatz and solve for it afterwards.

We exemplify that step with the supersymmetric limit of the CCLP black hole, see \cite{Chong:2005hr, Gutowski:2004ez, Cassani:2015upa}. In the orthotoric coordinates $\qty{y, x, \rho, \Phi, \Psi}$ it is the field configuration
\begin{align}\label{eq:susy-CCLP-ortho}
  \begin{aligned}
  \dd[]{s}^2 ={}& - (e^0)^2 + (e^1)^2 + (e^2)^2 + (e^3)^2 + (e^4)^2 \,, \\
  A ={}& - \frac{\sqrt{3}}{C_x} \qty(e^0 + \frac{\ell_5}{3} \ensuremath{\mathcal{P}} - f_0 \dd[]{y} + \dd[]{\lambda}) \,,
  \end{aligned}
\end{align}
where $f_0$ is a constant gauge shift, $\lambda(x, \Phi, \Psi)$ a gauge shift function, and the frame is
\begin{align}\label{eq:susy-frame}
  \begin{aligned}
  e^0 ={}& f \qty(\dd[]{y} + \varpi) \,, \\
  e^1 ={}& \ell_5 \sqrt{\frac{x - \rho}{f \ensuremath{\mathcal{F}}}} \dd[]{\rho} \,, \\
  e^2 ={}& \ell_5 \sqrt{\frac{\ensuremath{\mathcal{F}}}{f(x - \rho)}} \qty(\dd[]{\Phi} + x \dd[]{\Psi}) \,, \\
  e^3 ={}& - \ell_5 \sqrt{\frac{x - \rho}{f \ensuremath{\mathcal{G}}}} \dd[]{x} \,, \\
  e^4 ={}& \ell_5 \sqrt{\frac{\ensuremath{\mathcal{G}}}{f(x - \rho)}} \qty(\dd[]{\Phi} + \rho \dd[]{\Psi}) \,.
  \end{aligned}
\end{align}
Above $f = f(\rho,x)$, $\ensuremath{\mathcal{F}} = \ensuremath{\mathcal{F}}(\rho)$ and $\ensuremath{\mathcal{G}} = \ensuremath{\mathcal{G}}(x)$, while $\varpi$ and $\ensuremath{\mathcal{P}}$ are the one-forms
\begin{align}\label{eq:susy-varpi}
  \begin{aligned}
  \varpi ={}& \frac{\ell_5 \qty(\ensuremath{\mathcal{F}}''' + \ensuremath{\mathcal{G}}''')}{48 (x - \rho)^2} \qty[\qty(\ensuremath{\mathcal{F}} + (x - \rho) \qty(\frac{1}{2} \ensuremath{\mathcal{F}}' - \frac{1}{4} \ensuremath{\mathcal{F}}'''(\rho - m_1)^2)) \qty(\dd[]{\Phi} + x \dd[]{\Psi}) + \ensuremath{\mathcal{G}} \qty(\dd[]{\Phi} + \rho \dd[]{\Psi})] \\
            & - \frac{\ell_5 \ensuremath{\mathcal{F}}''' \ensuremath{\mathcal{G}}'''}{288} \qty((x + \rho) \dd[]{\Phi} + x \rho \dd[]{\Psi}) + \dd[]{\chi} \,, \\
  \ensuremath{\mathcal{P}} ={}& - \frac{1}{2(x - \rho)} \qty(\ensuremath{\mathcal{F}}' \qty(\dd[]{\Phi} + x \dd[]{\Psi}) + \ensuremath{\mathcal{G}}' \qty(\dd[]{\Phi} + \rho \dd[]{\Psi})) \,,
  \end{aligned}
\end{align}
with $\chi(x, \Phi, \Psi)$ a second gauge shift function. Supersymmetry fixes $f$ in terms of $\mathcal{F}$ and $\mathcal{G}$
\begin{align}\label{eq:f_relation}
  f ={}& \frac{24(x - \rho)}{\ensuremath{\mathcal{F}}'' + \ensuremath{\mathcal{G}}''} \,.
\end{align}
Furthermore $\mathcal{F}$ and $\mathcal{G}$ satisfy a complicated 6th order PDE. A known solution of it has both functions cubic
\begin{align}\label{eq:cubics}
  \ensuremath{\mathcal{G}}(x) ={}& g_0 (x^2 - 1) (x - g_1) \,, \qquad \ensuremath{\mathcal{F}}(\rho) = - \ensuremath{\mathcal{G}}(\rho) + m_0 (\rho - m_1)^3 \,,
\end{align}

Having recalled the physics background, we begin by including the needed definitions into \free{sol}
\begin{wlin}
IncludeTo[(!\free{sol}!)[\$chart], {
  <|Symbol -> (!\free{curved}!), Manifold -> (!\fre{\mathcal{M}}!),
   CNumbersOf -> {0, 1, 2, 3, 4},
   ScalarsOfChart -> {(!\free{y}!)[], (!\free{x}!)[], (!\fre{\rho}!)[], (!\fre{\Phi}!)[], (!\fre{\Psi}!)[]},
   ChartColor -> Red|>
  }]
IncludeTo[(!\free{sol}!)[\$form], {
  <|Symbol -> (!\fre{\varpi}!)[], Manifold -> (!\fre{\mathcal{M}}!), Deg -> 1, Symmetry -> None|>,
  <|Symbol -> (!\fre{\mathcal{P}}!)[], Manifold -> (!\fre{\mathcal{M}}!), Deg -> 1, Symmetry -> None|>
  }]
IncludeTo[(!\free{sol}!)[\$function], (<|Symbol -> (!\wlslot{\#}!)|> & /@ {(!\free{f}!)[(!\fre{\rho}!)[], (!\free{x}!)[]], (!\fre{\mathcal{G}}!)[(!\free{x}!)[]],
   (!\fre{\mathcal{F}}!)[(!\fre{\rho}!)[]], (!\fre{\lambda}!)[(!\free{x}!)[], (!\fre{\Phi}!)[], (!\fre{\Psi}!)[]], (!\fre{\chi}!)[(!\free{x}!)[], (!\fre{\Phi}!)[], (!\fre{\Psi}!)[]]})]
IncludeTo[(!\free{sol}!)[\$constant], (<|Symbol -> (!\wlslot{\#}!)|> & /@ {(!\free{g0}!), (!\free{g1}!), (!\free{m0}!), (!\free{m1}!), (!\free{f0}!)})]
IncludeTo[(!\free{sol}!)[\$assumption], <|Symbol -> (!\free{asymp}!), Expression -> {
    (!\free{l5}!) > 0,
    (!\fre{\mathcal{F}}!)[(!\fre{\rho}!)[]] > 0, (!\fre{\mathcal{G}}!)[(!\free{x}!)[]] > 0, (!\free{f}!)[(!\fre{\rho}!)[], (!\free{x}!)[]] > 0,
    Derivative[4][(!\fre{\mathcal{F}}!)][(!\fre{\rho}!)[]] == 0, Derivative[4][(!\fre{\mathcal{G}}!)][(!\free{x}!)[]] == 0,
    -1 < (!\free{x}!)[] < 1, (!\free{x}!)[] > (!\fre{\rho}!)[]
   }|>]
\end{wlin}
\begin{wlpanel}
(!\wlplain{Loaded:}!)
PDcurved, curved, y, x, @[Rho], @[CapitalPhi], @[CapitalPsi], dicurved, dcurved, TorsionPDcurved, ChristoffelPDcurved, RiemannPDcurved, RicciPDcurved, etaUpcurved, etaDowncurved, curved0, curved1, curved2, curved3, curved4, icurved, KoszulPDcurved, CovDPDcurved
(!\wlplain{Info[...]}!)
\end{wlpanel}
\begin{wlpanel}[1]
(!\wlplain{Loaded:}!)
T@[CurlyPi], @[CurlyPi], T@[ScriptCapitalP], @[ScriptCapitalP]
(!\wlplain{Info[...]}!)
\end{wlpanel}
\begin{wlpanel}[1]
(!\wlplain{Loaded:}!)
f, @[ScriptCapitalG], @[ScriptCapitalF], @[Lambda], @[Chi]
(!\wlplain{Info[...]}!)
\end{wlpanel}
\begin{wlpanel}[1]
(!\wlplain{Loaded:}!)
g0, g1, m0, m1, f0
(!\wlplain{Info[...]}!)
\end{wlpanel}
\begin{wlpanel}[1]
(!\wlplain{Loaded:}!)
\$Assumptions
(!\wlplain{Info[...]}!)
\end{wlpanel}
\noindent Notice the \code{[]} brackets next to the coordinate symbols; this is \xact's convention and signifies that the coordinates are really scalar functions of the chart \code{curved}. \code{sol[\$assumption]} is a special key: whatever \code{Expression} we include under it, and we may \sx{IncludeTo} it as many times as we like, becomes part of the kernel's \wlrefD{Assumptions}. For example, with the above, $\code{Sqrt[}\ensuremath{\mathcal{F}}\code{[]}^2\code{]}$ \code{Simplify}-es to $\ensuremath{\mathcal{F}}\code{[]}$. \code{sol[\$function]} also deserves a closer look. Its \code{Symbol} key carries more than a name. What is declared is the head \emph{together with its arguments}: \code{f[}$\rho$\code{[], x[]]}. The head becomes the \xact scalar function, defined through \wlref{DefScalarFunction}; the argument list becomes an abbreviation, so that from now on \code{f[]} evaluates to \code{f[}$\rho$\code{[], x[]]} and \code{Derivative[1, 0][f][]} evaluates to \code{Derivative[1, 0][f][}$\rho$\code{[], x[]]}\footnote{That abbreviation is why the frame and the metric below can be typed with bare \code{f[]}, $\ensuremath{\mathcal{F}}$\code{[]} and $\ensuremath{\mathcal{G}}$\code{[]} and still carry the right dependence}. We supply the three functions with their expressions
\begin{wlinm}
\noindent$\begin{aligned}
 &\free{sol}\code{[\$function, }\fre{\mathcal{G}}\code{, Expression]} =
   \free{g0}\;\big(\free{x}\code{[]}^{2} - 1\big)\,\big(\free{x}\code{[]} - \free{g1}\big) \\
 &\free{sol}\code{[\$function, }\fre{\mathcal{F}}\code{, Expression]} =
   -\fre{\mathcal{G}}\code{[}\fre{\rho}\code{[]]} + \free{m0}\;\big(\fre{\rho}\code{[]} - \free{m1}\big)^{3} \\
 &\free{sol}\code{[\$function, }\free{f}\code{, Expression]} =
   \frac{24\,\big(\free{x}\code{[]} - \fre{\rho}\code{[]}\big)}
        {\fre{\mathcal{F}}\code{''[}\fre{\rho}\code{[]]} + \fre{\mathcal{G}}\code{''[}\free{x}\code{[]]}}
\end{aligned}$
\end{wlinm}
\noindent Stored like that these expressions are inert: they are data in the record, not definitions in the kernel.
\sx{Gen}\footnote{Shortened from ``Generate'', as it is typed very often in a standard workflow.} is what turns them into genuine \wlref{Function} rules, for example
\begin{wlin}
Gen@(!\free{sol}!)[\$function, (!\free{f}!)]
\end{wlin}
\begin{wloutm}
\[
  \code{f} \to \code{Function}\Big[\{\code{xxx1}, \code{xxx2}\},\;
    \frac{24\,(-\code{xxx1} + \code{xxx2})}
         {\ensuremath{\mathcal{F}}''[\code{xxx1}] + \ensuremath{\mathcal{G}}''[\code{xxx2}]}\Big]
\]
\end{wloutm}
\noindent From here the solution data is typed almost verbatim as in the formulas above: the frame \eqref{eq:susy-frame}
\begin{wlinm}
\noindent$\begin{aligned}
 &\free{sol}\code{[\$frame, }\free{ee}\code{, Expression]} = \Bigg\{ \\
 &\quad \free{f}\code{[}\fre{\rho}\code{[], }\free{x}\code{[]]}\,
    \big(\code{Diff[}\free{y}\code{[]]} + \fre{\varpi}\code{[]}\big), \\
 &\quad \free{l5}\;\sqrt{\frac{\free{x}\code{[]} - \fre{\rho}\code{[]}}
    {\free{f}\code{[]}\;\fre{\mathcal{F}}\code{[]}}}\;\;\code{Diff[}\fre{\rho}\code{[]]}, \\
 &\quad \free{l5}\;\sqrt{\frac{\fre{\mathcal{F}}\code{[]}}
    {\free{f}\code{[]}\,\big(\free{x}\code{[]} - \fre{\rho}\code{[]}\big)}}\;\;
    \big(\code{Diff[}\fre{\Phi}\code{[]]} + \free{x}\code{[]}\wltimes\code{Diff[}\fre{\Psi}\code{[]]}\big), \\
 &\quad -\free{l5}\;\sqrt{\frac{\free{x}\code{[]} - \fre{\rho}\code{[]}}
    {\free{f}\code{[]}\;\fre{\mathcal{G}}\code{[]}}}\;\;\code{Diff[}\free{x}\code{[]]}, \\
 &\quad \free{l5}\;\sqrt{\frac{\fre{\mathcal{G}}\code{[]}}
    {\free{f}\code{[]}\,\big(\free{x}\code{[]} - \fre{\rho}\code{[]}\big)}}\;\;
    \big(\code{Diff[}\fre{\Phi}\code{[]]} + \fre{\rho}\code{[]}\wltimes\code{Diff[}\fre{\Psi}\code{[]]}\big) \\
 &\Bigg\}
\end{aligned}$
\end{wlinm}
\noindent and the metric \eqref{eq:susy-CCLP-ortho}
\begin{wlinm}
{\footnotesize
\noindent$\begin{aligned}
 &\free{sol}\code{[\$metric, }\free{gg}\code{, Expression]} = \\
 &\quad -\free{f}\code{[]}^{2}\,\big(\code{Diff[}\free{y}\code{[]]} + \fre{\varpi}\code{[]}\big)\otimes
    \big(\code{Diff[}\free{y}\code{[]]} + \fre{\varpi}\code{[]}\big) + \\
 &\quad \free{l5}^{2}\,\free{f}\code{[]}^{-1}\Bigg(
    \frac{\free{x}\code{[]} - \fre{\rho}\code{[]}}{\fre{\mathcal{F}}\code{[]}}\;
    \code{Diff[}\fre{\rho}\code{[]]}\otimes\code{Diff[}\fre{\rho}\code{[]]} \\
 &\quad\; + \frac{\fre{\mathcal{F}}\code{[]}}{\free{x}\code{[]} - \fre{\rho}\code{[]}}\;
    \big(\code{Diff[}\fre{\Phi}\code{[]]} + \free{x}\code{[]}\wltimes\code{Diff[}\fre{\Psi}\code{[]]}\big)\otimes
    \big(\code{Diff[}\fre{\Phi}\code{[]]} + \free{x}\code{[]}\wltimes\code{Diff[}\fre{\Psi}\code{[]]}\big) \\
 &\qquad + \frac{\free{x}\code{[]} - \fre{\rho}\code{[]}}{\fre{\mathcal{G}}\code{[]}}\;
    \code{Diff[}\free{x}\code{[]]}\otimes\code{Diff[}\free{x}\code{[]]} \\
 &\quad\; + \frac{\fre{\mathcal{G}}\code{[]}}{\free{x}\code{[]} - \fre{\rho}\code{[]}}\;
    \big(\code{Diff[}\fre{\Phi}\code{[]]} + \fre{\rho}\code{[]}\wltimes\code{Diff[}\fre{\Psi}\code{[]]}\big)\otimes
    \big(\code{Diff[}\fre{\Phi}\code{[]]} + \fre{\rho}\code{[]}\wltimes\code{Diff[}\fre{\Psi}\code{[]]}\big)\Bigg)
\end{aligned}$}
\end{wlinm}
\noindent Note that \solx registers the symbol \code{CircleTimes} ($\otimes$) as a graded product, within \xact's framework of products\footnote{See \code{Wedge} ($\wedge$) in \xterior and \code{CenterDot} ($\cdot$) in \fieldsx}, to represent the symmetric product of forms. The remaining input, $\varpi$ and $\ensuremath{\mathcal{P}}$ of \eqref{eq:susy-varpi} followed by $A$ of \eqref{eq:susy-CCLP-ortho} and $F = \dd A$, is typed the same way and we do not reproduce it here. 
Once the metric and all the forms are populated we register three rules that will be convenient during the \sx{Compute} phase
\begin{wlin}
IncludeTo[(!\free{sol}!)[\$rule], <|Symbol -> (!\free{func}!), Expression ->
  FoldedRule[Gen@(!\free{sol}!)[\$function, (!\wlslot{\#}!)] & /@ {(!\free{f}!)},
   Gen@(!\free{sol}!)[\$function, (!\wlslot{\#}!)] & /@ {(!\fre{\mathcal{F}}!)},
   Gen@(!\free{sol}!)[\$function, (!\wlslot{\#}!)] & /@ {(!\fre{\mathcal{G}}!)}]|>]
IncludeTo[(!\free{sol}!)[\$rule], <|Symbol -> (!\free{der3}!), Expression ->
  ((Derivative[3][(!\wlslot{\#}!)][] -> (Derivative[3][(!\wlslot{\#}!)][] /. Gen@(!\free{sol}!)[\$rule, (!\free{func}!)])) & /@ {(!\fre{\mathcal{F}}!), (!\fre{\mathcal{G}}!)})|>]
IncludeTo[(!\free{sol}!)[\$rule], <|Symbol -> \$form, Expression -> FoldedRule[
    Gen[(!\free{sol}!)[\$form, (!\wlslot{\#}!)]] & /@ {(!\free{F}!)},
    Gen[(!\free{sol}!)[\$form, (!\wlslot{\#}!)]] & /@ {(!\free{A}!)},
    Gen[(!\free{sol}!)[\$form, (!\wlslot{\#}!)]] & /@ {(!\fre{\varpi}!), (!\fre{\mathcal{P}}!)}
    ]|>]
\end{wlin}
\noindent Notice that some of them are layered, making use of \xact's \code{FoldedRule}\footnote{\code{FoldedRule} belongs to \xcore; see the \code{xCoreDoc} notebook shipped with \xact}. \free{der3} collapses the third derivatives of the two cubics, which are constants and therefore cheap in the upcoming computations. Thus far these rules are inertly stored. \sx{Gen} is overloaded on \code{\$rule}'s to call the actual replacement rules
\begin{wlin}
Gen@(!\free{sol}!)[\$rule, (!\free{func}!)]
Gen@(!\free{sol}!)[\$rule, (!\free{der3}!)]
Gen@(!\free{sol}!)[\$rule, \$form]
\end{wlin}
\begin{wloutm}
\begin{align*}
  \code{FoldedRule}\Big[
   &\big\{\code{f} \to \code{Function}\big[\{\code{xxx1}, \code{xxx2}\},\;
     \frac{24\,(-\code{xxx1} + \code{xxx2})}
          {\ensuremath{\mathcal{F}}''[\code{xxx1}] + \ensuremath{\mathcal{G}}''[\code{xxx2}]}\big]\big\}, \\
   &\big\{\ensuremath{\mathcal{F}} \to \code{Function}\big[\{\code{xxx1}\},\;
     \code{m0}\,(-\code{m1} + \code{xxx1})^{3} - \ensuremath{\mathcal{G}}[\code{xxx1}]\big]\big\}, \\
   &\big\{\ensuremath{\mathcal{G}} \to \code{Function}\big[\{\code{xxx1}\},\;
     \code{g0}\,(-\code{g1} + \code{xxx1})\,(-1 + \code{xxx1}^{2})\big]\big\}\Big]
\end{align*}
\end{wloutm}
\begin{wloutm}[1]
\[
  \big\{\ensuremath{\mathcal{F}}^{(3)}[\rho] \to -6\,\code{g0} + 6\,\code{m0},\;
        \ensuremath{\mathcal{G}}^{(3)}[\code{x}] \to 6\,\code{g0}\big\}
\]
\end{wloutm}
\begin{wloutm}[1]
{\small
\begin{align*}
  \code{FoldedRule}\Big[
   &\big\{\code{F} \to \code{d[A]}\big\}, \\
   &\Big\{\code{A} \to -\frac{\sqrt{3}}{\code{Cx}}\Big(-\code{f0}\,\code{d[y]}
      + \tfrac{\code{l5}\,\ensuremath{\mathcal{P}}}{3} + \code{f[}\rho\code{,x]}\,(\code{d[y]} + \varpi) \\
   &\qquad\qquad + \code{d[}\Psi\code{]}\,\lambda^{(0,0,1)}[\code{x},\Phi,\Psi]
      + \code{d[}\Phi\code{]}\,\lambda^{(0,1,0)}[\code{x},\Phi,\Psi]
      + \code{d[x]}\,\lambda^{(1,0,0)}[\code{x},\Phi,\Psi]\Big)\Big\}, \\
   &\Big\{\varpi \to
      -\frac{\code{l5}\,\big(\code{d[}\Psi\code{]}\,\code{x}\,\rho
        + \code{d[}\Phi\code{]}\,(\code{x} + \rho)\big)\,
        \ensuremath{\mathcal{F}}^{(3)}[\rho]\,\ensuremath{\mathcal{G}}^{(3)}[\code{x}]}{288} \\
   &\qquad
      + \frac{\code{l5}\,\big(\ensuremath{\mathcal{F}}^{(3)}[\rho] + \ensuremath{\mathcal{G}}^{(3)}[\code{x}]\big)}
             {48\,(\code{x} - \rho)^{2}}
        \Big(\ensuremath{\mathcal{G}}[\code{x}]\,(\code{d[}\Phi\code{]} + \code{d[}\Psi\code{]}\,\rho) \\
   &\qquad\qquad + (\code{d[}\Phi\code{]} + \code{d[}\Psi\code{]}\,\code{x})\,
      \big(\ensuremath{\mathcal{F}}[\rho] + (\code{x} - \rho)\big(\tfrac{\ensuremath{\mathcal{F}}'[\rho]}{2}
      - \tfrac{1}{4}(-\code{m1} + \rho)^{2}\,\ensuremath{\mathcal{F}}^{(3)}[\rho]\big)\big)\Big) \\
   &\qquad\qquad + \code{d[}\Psi\code{]}\,\chi^{(0,0,1)}[\code{x},\Phi,\Psi]
      + \code{d[}\Phi\code{]}\,\chi^{(0,1,0)}[\code{x},\Phi,\Psi]
      + \code{d[x]}\,\chi^{(1,0,0)}[\code{x},\Phi,\Psi], \\
   &\qquad \ensuremath{\mathcal{P}} \to -\frac{(\code{d[}\Phi\code{]} + \code{d[}\Psi\code{]}\,\code{x})\,\ensuremath{\mathcal{F}}'[\rho]
      + (\code{d[}\Phi\code{]} + \code{d[}\Psi\code{]}\,\rho)\,\ensuremath{\mathcal{G}}'[\code{x}]}
      {2\,(\code{x} - \rho)}\Big\}\Big]
\end{align*}}
\end{wloutm}
\noindent Notice that the gauge functions $\lambda$ and $\chi$ have been kept arbitrary. In \cref{sec:friends} that freedom will be spent to match the present solution with the supersymmetric limit of \sxdata{Curated/Sol__5dL__CCLP__Boyer-Lindquist}.

\subsection{Computing the tensor components of the solution}
\label{sec:compute}
So far nothing has been computed: what the record holds is a chart, a handful of functions and forms, and the expressions \eqref{eq:susy-CCLP-ortho}--\eqref{eq:cubics} written in them. Next we \sx{Compute} the tensor components out of these definitions.

\paragraph{The chain} Every entry that \sx{Compute} can produce is reached through a \code{Chain}: an ordered list saying where the components come from. A chain entry has one of two shapes. An \emph{array} entry: \wlref{HoldForm}\code{[}\sxarg{expr}\code{]} \ensuremath{\rightarrow} \code{\{}\sxarg{slots}\code{\}}, holds an expression that \wlref{ReleaseHold}-s into an explicit array, which is then stored into the named index placement. A \emph{slot} entry: \code{\{}\sxarg{from}\code{\}} \ensuremath{\rightarrow} \code{\{}\sxarg{to}\code{\}}, will compute the left-hand side by raising or lowering the right-hand side with the metrics of the bundles involved.
For example, the \textit{set of computation instructions} for the metric, which we call \code{Routine}, is
\begin{wlin}
(!\free{sol}!)[\$metric, (!\free{gg}!), \$auto, (!\free{gg}!), Routine, Chain] = {
  HoldForm[MakeArray[\$self[\$metric, (!\free{gg}!), Expression] /. Gen@\$self[\$rule, \$form]]] -> {{-1, -1}},
  HoldForm[Inverse@ToArray@(!\free{gg}!)[-(!\fre{\curly{m}}!), -(!\fre{\curly{n}}!)]] -> {{1, 1}}
  }
\end{wlin}
\noindent \code{\$self} is \solx's way to emulate Python's \code{self}. It refers to the parent association, whatever that association happens to be called. For longer chains there is the shorthand syntax \sx{RulesToChain} which expands an arrow sequence into the successive pairs; e.g. \code{RulesToChain[a}\;\ensuremath{\rightarrow}\;\code{b}\;\ensuremath{\rightarrow}\;\code{c]} gives \code{\{a}\;\ensuremath{\rightarrow}\;\code{b, b}\;\ensuremath{\rightarrow}\;\code{c\}}. Using it, the Christoffel symbols of the Levi-Civita connection have the \code{Chain}
\begin{wlin}
(!\free{sol}!)[\$metric, (!\free{gg}!), \$auto, (!\free{Christoffelcd}!), Routine, Chain] = RulesToChain @@ {
   HoldForm[ToArray@ChristoffelToGradMetric@(!\free{Christoffelcd}!)[-(!\fre{\curly{m}}!), -(!\fre{\curly{n}}!), -(!\fre{\curly{o}}!)]] ->
    Permutations[{-1, -1, -1}] -> Permutations[{1, -1, -1}]
   }
\end{wlin}
\noindent a.k.a. build the all-lower placement from $\partial g$, then raise one index. The vielbein, which lives in two bundles at once, is a three-step chain
\begin{wlin}
(!\free{sol}!)[\$frame, (!\free{ee}!), \$auto, (!\free{ee}!), Routine, Chain] = RulesToChain @@ {
   HoldForm[Transpose[MakeArray[(!\wlslot{\#}!)] & /@ (\$self[\$frame, (!\free{ee}!), Expression] /. Gen@\$self[\$rule, \$form])]] ->
    {{-1, 1}} -> {{1, 1}, {-1, -1}} -> {{1, -1}}
   }
\end{wlin}
\noindent The spin connection is a single array entry, written straight from its definition in terms of the Christoffel symbols and the vielbein
\begin{wlin}
(!\free{sol}!)[\$spinConnection, (!\fre{\omega\omega}!), \$auto, (!\fre{\omega\omega}!), Routine, Chain] = {
  HoldForm[ToArray@((!\free{Christoffelcd}!)[(!\fre{\curly{o}}!), -(!\fre{\curly{n}}!), -(!\fre{\curly{m}}!)] (!\free{ee}!)[(!\fre{\curly{n}}!), -(!\fre{\mathfrak{c}}!)] (!\free{ee}!)[-(!\fre{\curly{o}}!), -(!\fre{\mathfrak{b}}!)]
     - (!\free{ee}!)[(!\fre{\curly{p}}!), -(!\fre{\mathfrak{c}}!)] PD[-(!\fre{\curly{m}}!)][(!\free{ee}!)[-(!\fre{\curly{p}}!), -(!\fre{\mathfrak{b}}!)]])] -> {{-1, -1, -1}}
  }
\end{wlin}
\noindent It is worth being plain about what a slot specification buys, because it is easy to write more of them than one needs. A list of $\pm 1$ says where each index sits: \code{\{-1, -1, -1\}} is ``all three downstairs'', \code{\{1, -1, -1\}} is ``the first one upstairs''. Getting from the first to the second costs a contraction with the metric. That is real work, and it is why it is a rung of its own. Getting from \code{\{1, -1, -1\}} to \code{\{-1, 1, -1\}} might cost less depending on the symmetries of the tensor. \sx{Compute} fills all such rearrangements together, so the three placements that \code{Permutations[\{1, -1, -1\}]} lists above are one rung of the ladder and not three. The rule of thumb: a different \emph{number} of raised indices needs its own chain entry, a different \emph{order} of the same indices does not.
Within a placement the same economy applies. \sx{Compute} evaluates only the components the tensor's symmetries leave independent, and fills the rest from those, signs included\footnote{The all-lower Christoffel symbols below are symmetric in their last two indices, so 75 of their 125 components are evaluated and the other 50 are read off.}.

\paragraph{What is done to each component} Every value \sx{Compute} produces is passed through the functions of the \code{Apply} option, whose default is the global \sxD{Apply}: a list keyed by \wlref{Map} and \wlref{ParallelMap}, applied in that order: the first serially, the second distributed over the subkernels.
For this record we set
\begin{wlin}
\$Apply = {
  Map -> ((!\wlslot{\#}!) /. Gen@(!\free{sol}!)[\$rule, (!\free{der3}!)] &),
  ParallelMap -> (Simplify[Together[(!\wlslot{\#}!)]] &)
  }
\end{wlin}
\noindent so that the third derivatives of the cubics collapse to constants before anything expensive happens, and the simplification runs in parallel. The split is not a matter of taste. A function handed to \code{ParallelMap} must be built from \code{System`} symbols alone\footnote{This is a present \mma limitation to which we have not yet found a workaround.}; \code{Gen} and \free{sol} are not, which is why that substitution sits in the \code{Map} slot. \code{Simplify[Together[\#]] \&} qualifies, and the assumptions of \code{sol[\$assumption]} are shipped with it, which is what lets \wlref{Simplify} combine the radicals of the frame.

\paragraph{Computing} With the chains stored and \sxD{Apply} set, we are finally ready to \sx{Compute}
\wlapply{\#1 /. Gen[sol[\$rule, der3]]~\&}{Simplify[Together[\#1]]~\&}%
\begin{wlin}
Compute@(!\free{sol}!)[\$metric, (!\free{gg}!), \$auto, (!\wlslot{\#}!)] & /@ {(!\free{gg}!), (!\free{Detgg}!), (!\free{Christoffelcd}!), (!\free{Riccicd}!), (!\free{RicciScalarcd}!)};
\end{wlin}
\begin{wlecho}
\wlecholine{$\code{gg}_{\ichart{\curly{m}}\ichart{\curly{n}}} = {}$%
  \code{MakeArray[\$self[\$metric, gg, Expression]}\allowbreak\code{ /. Gen[\$self[\$rule, \$form]]]}}{0min 11s}
\wlecholines{$\code{gg}^{\ichart{\curly{m}}\ichart{\curly{n}}} =
  \code{gg}_{\ichart{\curly{m}}\ichart{\curly{n}}}^{\,-1}$}{0min 2s}
\wlecholines{$\wldensityupp{\code{g}}\code{g} =
  \code{Det}\big[\code{gg}_{\ichart{\curly{m}}\ichart{\curly{n}}}\big]$}{0min 0s}
\wlecholines{$\Gamma[\nabla]_{\ichart{\curly{m}}\ichart{\curly{n}}\ichart{\curly{o}}} = \tfrac{1}{2}\big(
  {-\partial_{\ichart{\curly{m}}}}\code{gg}_{\ichart{\curly{n}}\ichart{\curly{o}}}
  + \partial_{\ichart{\curly{n}}}\code{gg}_{\ichart{\curly{o}}\ichart{\curly{m}}}
  + \partial_{\ichart{\curly{o}}}\code{gg}_{\ichart{\curly{n}}\ichart{\curly{m}}}\big)$}{0min 14s}
\wlecholines{$\Gamma[\nabla]\indices{^{\ichart{\curly{m}}}_{\ichart{\curly{n}}\ichart{\curly{o}}}},\;
  \Gamma[\nabla]\indices{_{\ichart{\curly{m}}}^{\ichart{\curly{n}}}_{\ichart{\curly{o}}}},\;
  \Gamma[\nabla]\indices{_{\ichart{\curly{m}}\ichart{\curly{n}}}^{\ichart{\curly{o}}}}$\;\code{obtained from}\;
  $\Gamma[\nabla]_{\ichart{\curly{m}}\ichart{\curly{n}}\ichart{\curly{o}}}$\;\code{using metrics}\;
  $\mathbb{T}\ensuremath{\mathcal{M}} \rightarrow \code{gg}$}{1min 44s}
\wlecholines{$\code{R}[\nabla]_{\ichart{\curly{m}}\ichart{\curly{n}}} =
  -\Gamma[\nabla]\indices{^{\ichart{\curly{o}}}_{\ichart{\curly{m}}\ichart{\curly{p}}}}
   \Gamma[\nabla]\indices{^{\ichart{\curly{p}}}_{\ichart{\curly{o}}\ichart{\curly{n}}}}
  + \Gamma[\nabla]\indices{^{\ichart{\curly{o}}}_{\ichart{\curly{m}}\ichart{\curly{n}}}}
    \Gamma[\nabla]\indices{^{\ichart{\curly{p}}}_{\ichart{\curly{p}}\ichart{\curly{o}}}}
  - \partial_{\ichart{\curly{m}}}\Gamma[\nabla]\indices{^{\ichart{\curly{o}}}_{\ichart{\curly{o}}\ichart{\curly{n}}}}
  + \partial_{\ichart{\curly{o}}}\Gamma[\nabla]\indices{^{\ichart{\curly{o}}}_{\ichart{\curly{m}}\ichart{\curly{n}}}}$}{3min 33s}
\wlecholines{$\code{R}[\nabla]\indices{^{\ichart{\curly{m}}}_{\ichart{\curly{n}}}},\;
  \code{R}[\nabla]\indices{_{\ichart{\curly{m}}}^{\ichart{\curly{n}}}}$\;\code{obtained from}\;
  $\code{R}[\nabla]_{\ichart{\curly{m}}\ichart{\curly{n}}}$\;\code{using metrics}\;
  $\mathbb{T}\ensuremath{\mathcal{M}} \rightarrow \code{gg}$}{2min 12s}
\wlecholines{$\code{R}[\nabla] = \code{Tr}\big[\code{R}[\nabla]\indices{^{\ichart{\curly{m}}}_{\ichart{\curly{n}}}}\big]$}{0min 2s}
\end{wlecho}
\noindent \solx echoes each step in the chain: how it was computed and how long it took. The frame goes the same way. Its flat metric is entered as a literal and inverted; the vielbein is built from the frame expression \eqref{eq:susy-frame} and then walked through its three remaining placements, with the \code{flat} and \code{curved} indices being raised and lowered with $\eta\eta$ and \code{gg} respectively
\begin{wlin}
Compute@(!\free{sol}!)[\$frame, (!\free{ee}!), \$auto, (!\wlslot{\#}!)] & /@ {(!\fre{\eta\eta}!), (!\free{ee}!)};
\end{wlin}
\begin{wlecho}
\wlecholines{$\eta\eta_{\iflat{\mathfrak{a}}\iflat{\mathfrak{b}}} =
  \code{DiagonalMatrix[\{-1, 1, 1, 1, 1\}]}$}{0min 0s}
\wlecholines{$\eta\eta^{\iflat{\mathfrak{a}}\iflat{\mathfrak{b}}} =
  \eta\eta_{\iflat{\mathfrak{a}}\iflat{\mathfrak{b}}}^{\,-1}$}{0min 0s}
\wlecholines{$\code{ee}\indices{_{\ichart{\curly{m}}}^{\iflat{\mathfrak{b}}}} = {}$%
  \code{((MakeArray[\#1] \&) /@ (\$self[\$frame, ee, Expression]}\allowbreak%
  \code{ /. Gen[\$self[\$rule, \$form]]))}$^{\code{T}}$}{0min 0s}
\wlecholines{$\code{ee}^{\ichart{\curly{m}}\iflat{\mathfrak{b}}},\;
  \code{ee}_{\ichart{\curly{m}}\iflat{\mathfrak{b}}}$\;\code{obtained from}\;
  $\code{ee}\indices{_{\ichart{\curly{m}}}^{\iflat{\mathfrak{b}}}}$\;\code{using metrics}\;
  $\mathbb{T}\ensuremath{\mathcal{M}} \rightarrow \code{gg},\;
   \mathbb{F}\ensuremath{\mathcal{M}} \rightarrow \eta\eta$}{0min 1s}
\wlecholines{$\code{ee}\indices{^{\ichart{\curly{m}}}_{\iflat{\mathfrak{b}}}}$\;\code{obtained from}\;
  $\code{ee}^{\ichart{\curly{m}}\iflat{\mathfrak{b}}},\;\code{ee}_{\ichart{\curly{m}}\iflat{\mathfrak{b}}}$\;\code{using metrics}\;
  $\mathbb{T}\ensuremath{\mathcal{M}} \rightarrow \code{gg},\;
   \mathbb{F}\ensuremath{\mathcal{M}} \rightarrow \eta\eta$}{0min 0s}
\end{wlecho}
\noindent The spin connection is one entry and one echo:
\begin{wlin}
Compute@(!\free{sol}!)[\$spinConnection, (!\fre{\omega\omega}!), \$auto, (!\wlslot{\#}!)] & /@ {(!\fre{\omega\omega}!)};
\end{wlin}
\begin{wlecho}
\wlecholines{$\omega\omega_{\ichart{\curly{m}}\iflat{\mathfrak{b}}\iflat{\mathfrak{c}}} =
  \Gamma[\nabla]\indices{^{\ichart{\curly{o}}}_{\ichart{\curly{n}}\ichart{\curly{m}}}}\,
  \code{ee}\indices{^{\ichart{\curly{n}}}_{\iflat{\mathfrak{c}}}}\,
  \code{ee}_{\ichart{\curly{o}}\iflat{\mathfrak{b}}}
  - \code{ee}\indices{^{\ichart{\curly{p}}}_{\iflat{\mathfrak{c}}}}\,
    \partial_{\ichart{\curly{m}}}\code{ee}_{\ichart{\curly{p}}\iflat{\mathfrak{b}}}$}{0min 24s}
\end{wlecho}
\noindent Every index in an echo wears the colour of the basis it sits in: \basisswatch{sxchart} for the chart \code{curved}, \basisswatch{sxflat} for \code{flat}, \basisswatch{sxspin} for \code{spin}. Thus, one can read at a glance that $\omega\omega$ has one curved leg and two frame legs. The remaining sectors are computed analogously: \code{Compute@sol[\$tensor, \#, \$auto, \#] \& /@ \{TA, TF\}} builds the tensor partners of the gauge potential and its field strength from \code{sol[\$form, A, Expression]} and \code{sol[\$form, F, Expression]}, and \code{Compute@sol[\$spinStructure, gg, \$auto, \#] \& /@ \{...\}} does the same for the rank-one, -two and -three $\gamma$-matrices in both the flat and the curved bases. 

\subsection{Verifying the solution}
\label{sec:verify}
Having computed all the desired tensor values it is now time to check if the equations of motion and the Killing spinor equation hold. Since the solution is built on top of its parent theory (which stored these equations) we simply call them, act with \sx{ToArray} to convert them to arrays and then \wlref{Simplify}

\begin{wlin}
(!\free{\ensuremath{\ensuremath{\mathcal{E}}}gg}!)[-(!\fre{\curly{m}}!), -(!\fre{\curly{n}}!)]
(ToArray[
AbsoluteTiming[ParallelMap[Simplify, 
\end{wlin}
\begin{wloutm}
\[
  \frac{4\,\code{gg}_{\curly{m}\curly{n}}}{\code{l5}^2}
  + \code{R}[\nabla]_{\curly{m}\curly{n}}
  - \frac{\code{Cx}^2}{2}\,
    \code{TF}\indices{_{\curly{m}}^{\curly{o}}}\,\code{TF}_{\curly{n}\curly{o}}
  + \frac{\code{Cx}^2}{12}\,
    \code{gg}_{\curly{m}\curly{n}}\,
    \code{TF}_{\curly{o}\curly{p}}\,\code{TF}^{\curly{o}\curly{p}}
\]
\end{wloutm}
\begin{wlout}[2]
{53.878, {0, 0, 0, 0, 0, 0, 0, 0, 0, 0, 0, 0, 0, 0, 0, 0}}
\end{wlout}
\noindent The first line only recalls the stored equation. \sx{ToArray} takes it to bases, traces the dummies and inserts the component values computed in the previous subsection; \wlref{Flatten}, \wlref{DeleteCases} and \wlref{DeleteDuplicates} reduce the resulting $5 \times 5$ array to its $16$ distinct nonzero components. The \free{func} rule expands $f$, $\ensuremath{\mathcal{F}}$ and $\ensuremath{\mathcal{G}}$. That produces a set of long expressions, omitted here, each of which simplifies to $0$.

The Maxwell equation goes the same way, but it needs one more line:

\begin{wlin}
(!\free{\ensuremath{\ensuremath{\mathcal{E}}}TA}!)[-(!\fre{\curly{m}}!)]
(ToArray[
AbsoluteTiming[ParallelMap[Simplify, 
\end{wlin}
\begin{wloutm}
\[
  -\frac{\code{Cx}\,
    \epsilon\code{gg}_{\curly{m}\curly{n}\curly{o}\curly{p}\curly{q}}\,
    \code{TF}^{\curly{n}\curly{o}}\,\code{TF}^{\curly{p}\curly{q}}}{4\sqrt{3}}
  + \nabla_{\curly{n}}\code{TF}\indices{_{\curly{m}}^{\curly{n}}}
\]
\end{wloutm}
\begin{wloutm}[1]
\[
  \Gamma[\nabla]\indices{^{\curly{n}}_{\curly{n}\curly{o}}}\,
     \code{TF}\indices{_{\curly{m}}^{\curly{o}}}
   - \Gamma[\nabla]\indices{^{\curly{o}}_{\curly{n}\curly{m}}}\,
     \code{TF}\indices{_{\curly{o}}^{\curly{n}}}
   - \frac{\code{Cx}\,\sqrt{-\wldensityupp{\code{g}}\code{g}}\;
     \wldensity{\eta}_{\curly{m}\curly{n}\curly{o}\curly{p}\curly{q}}\,
     \code{TF}^{\curly{n}\curly{o}}\,\code{TF}^{\curly{p}\curly{q}}}{4\sqrt{3}}
   + \partial_{\curly{n}}\code{TF}\indices{_{\curly{m}}^{\curly{n}}}
\]
\end{wloutm}
\begin{wlout}[2]
{15.1216, {0, 0, 0, 0, 0}}
\end{wlout}
\noindent The extra line trades the two objects whose components we have not actually computed. \xcoba's \code{epsilonToetaDown[gg, curved]} rewrites the Levi-Civita tensor $\epsilon\code{gg}$ as the \code{etaDown} tensor of the chart \code{curved}\footnote{the antisymmetric symbol of the basis covectors, a density of weight $-1$ with components $0$ and $\pm 1$} times $\sqrt{-\wldensityupp{\code{g}}\code{g}}$. The tildes are \xcoba's way of signifying that the object in question is a tensor density\footnote{See the \code{xCobaDoc} notebook shipped with \xact}. \xtensor's \code{CovDToChristoffel} rewrites $\nabla$ as the chart's partial derivative plus Christoffel tensors. \wlref{PowerExpand}, which 
behaves as $\sqrt{ab} = \sqrt{a}\,\sqrt{b}$, is safe to use in this case because
at the very beginning we arranged positive factors under the square roots with \code{sol[\$assumption]}.

The Killing spinor equation is checked the same way, once the spinor is in the record. It is not derived but posed, and it enters through \sx{Compute} like any other component value:

\begin{wlinm}
\noindent$\begin{aligned}
 &\code{Compute}\Bigl[\free{sol}\code{[\$spinor, }\fre{\epsilon\epsilon}\code{, \$auto, }\fre{\epsilon\epsilon}\code{]}, \\[2pt]
 &\qquad \code{Chain} \to \Bigl\{\code{HoldForm}\Bigl[
     \eu^{\frac{3\iu}{2\,\free{l5}}\left(\free{f0}\,\free{y}\code{[]} - \fre{\lambda}\code{[]}\right)}\,
     \free{f}\code{[]}^{1/2}\,\{1, 1, -1, -1\}\Bigr] \to \{\{-1\}\}\Bigr\}, \\[2pt]
 &\qquad \code{Apply} \to \{\code{Map} \to \code{Identity}\}\Bigr]
\end{aligned}$
\end{wlinm}
\begin{wlecho}
\[
  \wlchev\;\code{Applied}\;\wlcompute{Identity}\;\code{to}\;\;
  \epsilon\epsilon_{\ispin{\mathbb{A}}} = \Bigl\{
     \eu^{\frac{3\iu\left(\code{f0}\,\code{y} - \lambda[\code{x},\Phi,\Psi]\right)}{2\,\code{l5}}}
       \sqrt{\code{f[}\rho\code{,x]}},\;
     \eu^{\frac{3\iu\left(\code{f0}\,\code{y} - \lambda[\code{x},\Phi,\Psi]\right)}{2\,\code{l5}}}
       \sqrt{\code{f[}\rho\code{,x]}},
\]
\[
     -\eu^{\frac{3\iu\left(\code{f0}\,\code{y} - \lambda[\code{x},\Phi,\Psi]\right)}{2\,\code{l5}}}
       \sqrt{\code{f[}\rho\code{,x]}},\;
     -\eu^{\frac{3\iu\left(\code{f0}\,\code{y} - \lambda[\code{x},\Phi,\Psi]\right)}{2\,\code{l5}}}
       \sqrt{\code{f[}\rho\code{,x]}}\Bigr\}\;\;
  \code{in}\;\wltime{0min 0s}
\]
\end{wlecho}
\noindent Nothing is really computed here: the array is simply hooked to the \code{TensorValues} of the spinor $\epsilon\epsilon$, so that it can be recalled with \sx{ToArray}.
After this preparation we are ready to check the Killing spinor equation

\begin{wlin}
(!\fre{\delta\psi\psi}!)[-(!\fre{\curly{m}}!), -(!\fre{\mathbb{A}}!)]
AbsoluteTiming[(ToArray[
\end{wlin}
\begin{wloutm}
\begin{align*}
  &-\frac{\iu\sqrt{3}\,\code{Cx}\,\code{TA}_{\curly{m}}\,
      \epsilon\epsilon_{\mathbb{A}}}{2\,\code{l5}}
   -\frac{\gamma[\code{gg}]\indices{_{\curly{m}}_{\mathbb{A}}^{\mathbb{B}}}\,
      \epsilon\epsilon_{\mathbb{B}}}{2\,\code{l5}}
   +\frac{\iu\,\code{Cx}\,
      \gamma[\code{gg}]\indices{^{\curly{n}}_{\mathbb{A}}^{\mathbb{B}}}\,
      \code{TF}_{\curly{m}\curly{n}}\,
      \epsilon\epsilon_{\mathbb{B}}}{2\sqrt{3}} \\[2pt]
  &\quad-\frac{\iu\,\code{Cx}\,
      \gamma[\code{gg}]\indices{_{\curly{m}}_{\curly{o}}_{\curly{n}}_{\mathbb{A}}^{\mathbb{B}}}\,
      \code{TF}^{\curly{o}\curly{n}}\,
      \epsilon\epsilon_{\mathbb{B}}}{8\sqrt{3}}
   +\frac{1}{4}\,
      \gamma[\eta\eta]\indices{^{\mathfrak{b}}^{\mathfrak{a}}_{\mathbb{A}}^{\mathbb{B}}}\,
      \epsilon\epsilon_{\mathbb{B}}\,
      \omega\omega_{\curly{m}\mathfrak{b}\mathfrak{a}}
   +\partial_{\curly{m}}\epsilon\epsilon_{\mathbb{A}}
\end{align*}
\end{wloutm}
\begin{wlout}[1]
{74.2331, {0, 0, 0, 0, 0, 0, 0, 0, 0, 0, 0, 0, 0, 0, 0, 0, 0, 0, 0, 0}}
\end{wlout}
\noindent With that, the solution is verified against the equations of motion and the Killing spinor equation of its parent theory.

\subsection{The solution can now play with its friends}
\label{sec:friends}
A verified record is worth keeping. \sx{SaveData} writes it, together with the notebook it was built in, into \code{\$DataDirectory/\$Alias/}\sxarg{name}\code{/}. The \sxD{DataDirectory} and the \sxD{Alias} are configured through the \sx{Welcome}\code{[]} wizard, or later with \sx{SetDataDirectory} and \sx{SetAlias}, and are recorded in the user configuration file, so they persist across sessions. \sxarg{name} is determined from the association itself
\begin{wlin}
(!\free{sol}!)[\$manifold, (!\fre{\mathcal{M}}!), Name] = "CCLP-susy";
(!\free{sol}!)[\$chart, (!\free{curved}!), Name] = "Orthotoric";
SaveData@(!\free{sol}!)
\end{wlin}
\noindent \sx{SaveData} echoes where it put the entry: \code{Saved}\;\sxarg{alias}\code{/}\sxarg{name}\;\code{under}\;\sxarg{directory}. What lands on disk is a folder named \code{\detokenize{Sol__5dL__CCLP-susy__Orthotoric}}, containing two files that carry that same name: \sxarg{name}\code{.nb}, the notebook from which we last called \sx{SaveData}, and \sxarg{name}\code{.m}, a compressed record of the entire association.
From now on the solution can be recalled with \sx{GetData} and \sx{Load}, which rebuilds the whole \xact environment together with every component computed in \cref{sec:compute}. \solx makes one simplifying assumption: one record is to be loaded at a time\footnote{We have not yet stumbled on a convincing argument for extending this to multiple entries loaded simultaneously.}. Thus, there is a single \xact environment in a kernel, and it belongs to whoever called \sx{Load}. Any number of further records may be \emph{fetched}. A fetched record is an ordinary association: its stored expressions can be read out and used, but its own symbols are inert. Below we give three physically relevant examples of how a stored solution can ``interact with its neighbours''.

\paragraph{The supersymmetric black hole is a limit of the non-supersymmetric black hole} The corpus already contains the general, non-super\-symmetric CCLP black hole in Boyer--Lindquist coordinates, \sxdata{Curated/Sol__5dL__CCLP__Boyer-Lindquist}, built on top of the same parent theory. In a fresh kernel we fetch our record, load it, and fetch that one
\begin{wlin}
(!\free{sol}!) = GetData["Curated/Sol__5dL__CCLP-susy__Orthotoric"];
Load@(!\free{sol}!)
(!\free{cclp}!) = GetData["Curated/Sol__5dL__CCLP__Boyer-Lindquist"];
\end{wlin}
\noindent The supersymmetric solution should be the limit of the general one, and the statement has two parts: a change of coordinates, and a restriction of parameters. The change of coordinates is entered as an ordinary list of rules
\begin{wlinm}
{\small
\noindent$\begin{aligned}
 &\free{ct} = \Bigg\{ \\
 &\quad \free{t}\code{[]} \rightarrow \free{y}\code{[]}, \\
 &\quad \fre{\theta}\code{[]} \rightarrow \sqrt{\frac{1 - \free{x}\code{[]}}{2}}, \\
 &\quad \free{r}\code{[]} \rightarrow \sqrt{\frac{1}{2}\,\big(\free{a}^{2} - \free{b}^{2}\big)\,\free{mt}\;\fre{\rho}\code{[]}
    + \frac{1}{\free{g}}\,\big((\free{a} + \free{b})\,\free{mt} + \free{a} + \free{b} + \free{a}\,\free{b}\,\free{g}\big)
    + \frac{1}{2}\,(\free{a} + \free{b})^{2}\,\free{mt}}, \\
 &\quad \fre{\phi 1}\code{[]} \rightarrow \free{g}\,\free{y}\code{[]} - 4\,
    \frac{1 - \free{a}^{2}\,\free{g}^{2}}{\big(\free{a}^{2} - \free{b}^{2}\big)\,\free{g}^{2}\,\free{mt}}\;
    \big(\fre{\Phi}\code{[]} - \fre{\Psi}\code{[]}\big), \\
 &\quad \fre{\phi 2}\code{[]} \rightarrow \free{g}\,\free{y}\code{[]} - 4\,
    \frac{1 - \free{b}^{2}\,\free{g}^{2}}{\big(\free{a}^{2} - \free{b}^{2}\big)\,\free{g}^{2}\,\free{mt}}\;
    \big(\fre{\Phi}\code{[]} + \fre{\Psi}\code{[]}\big) \\
 &\Bigg\}
\end{aligned}$}
\end{wlinm}
\noindent The parameter map is then obtained by matching polynomials. The change of variables above and the identifications below are those of \cite{Cassani:2015upa}, and the two cells that follow are the elementary algebra that pins the four constants of \eqref{eq:cubics} onto parameters of \sxdata{Curated/Sol__5dL__CCLP__Boyer-Lindquist}:
\begin{wlinm}
{\small
\noindent$\begin{aligned}
 &\fre{\mathcal{G}}\code{[] /. Gen@}\free{sol}\code{[\$rule, }\free{func}\code{];} \\
 &-\frac{4}{\big(\free{a}^{2} - \free{b}^{2}\big)\,\free{g}^{2}\,\free{mt}}\;
   \big(1 - \free{x}\code{[]}^{2}\big)\,
   \Big(\big(1 - \free{a}^{2}\,\free{g}^{2}\big)\big(1 + \free{x}\code{[]}\big)
      + \big(1 - \free{b}^{2}\,\free{g}^{2}\big)\big(1 - \free{x}\code{[]}\big)\Big)\code{;} \\
 &\code{CoefficientArrays[\%\% - \%, }\free{x}\code{[]] // Normal // Flatten // Simplify;} \\
 &\free{g01} = \code{Solve[\% == 0, \{}\free{g0}\code{, }\free{g1}\code{\}] // First}
\end{aligned}$}
\end{wlinm}
\begin{wloutm}[3]
\[
  \Big\{\code{g0} \to -\frac{4}{\code{mt}},\;\;
        \code{g1} \to \frac{2 - \code{a}^2\code{g}^2 - \code{b}^2\code{g}^2}
                          {\code{a}^2\code{g}^2 - \code{b}^2\code{g}^2}\Big\}
\]
\end{wloutm}
\noindent and, in the same way:
\begin{wlinm}
{\small
\noindent$\begin{aligned}
 &\fre{\mathcal{F}}\code{[] /. Gen@}\free{sol}\code{[\$rule, }\free{func}\code{];} \\
 &-\fre{\mathcal{G}}\code{[}\fre{\rho}\code{[]]} - 4\,\frac{1 + \free{mt}}{\free{mt}}\;
   \Bigg(\frac{\big(2 + \free{a}\,\free{g} + \free{b}\,\free{g}\big)}
              {\big(\free{a} - \free{b}\big)\,\free{g}} + \fre{\rho}\code{[]}\Bigg)^{\!3}
   \code{ /. Gen@}\free{sol}\code{[\$rule, }\free{func}\code{];} \\
 &\code{CoefficientArrays[\%\% - \%, }\fre{\rho}\code{[]] // Normal // Flatten // Simplify;} \\
 &\free{m01} = \code{Solve[\% == 0, \{}\free{m0}\code{, }\free{m1}\code{\}] // First}
\end{aligned}$}
\end{wlinm}
\begin{wloutm}[3]
\[
  \Big\{\code{m0} \to -\frac{4\,(1 + \code{mt})}{\code{mt}},\;\;
        \code{m1} \to \frac{-2 - \code{a}\,\code{g} - \code{b}\,\code{g}}
                          {\code{a}\,\code{g} - \code{b}\,\code{g}}\Big\}\,.
\]
\end{wloutm}
\noindent Everything is now in place to compare the two metrics. The fetched record's expression is pulled through the coordinate map and its own stored rules\footnote{Note that the potent combination of \code{Diff} and $\otimes$ does most of the heavy lifting automatically.}; ours is expanded with its form rule; the difference is turned into an array:
\begin{wlinm}
{\small
\noindent$\begin{aligned}
 &\free{cclp}\code{[\$metric, }\free{gg}\code{, Expression] /. }\free{ct}\code{ /. Gen@}\free{cclp}\code{[\$rule, }\free{func}\code{]} \\
 &\qquad \code{ /. Gen@}\free{cclp}\code{[\$rule, }\free{const}\code{];} \\
 &\free{sol}\code{[\$metric, }\free{gg}\code{, Expression] /. Gen@}\free{sol}\code{[\$rule, \$form];} \\
 &\code{(MakeArray[\%\% - \%] /. Gen@}\free{sol}\code{[\$rule, }\free{func}\code{]} \\
 &\qquad \code{// DeleteDuplicates // DeleteCases[0]);} \\
 &\Bigg(\code{\% /. }\fre{\chi} \rightarrow \code{Function}\Big[\{\wlslot{x}, \wlslot{\ensuremath{\Phi}}, \wlslot{\ensuremath{\Psi}}\},\;
    -\frac{2\,\wlslot{\ensuremath{\Psi}}}{\free{g}\;\free{mt}}\Big] \\
 &\qquad \code{ /. }\free{q} \rightarrow \frac{\free{m}}{1 + \free{a}\,\free{g} + \free{b}\,\free{g}}
    \code{ /. }\free{g01}\code{ /. }\free{m01}\code{ /.} \\
 &\qquad \free{mt} \rightarrow \frac{\free{m}\,\free{g}}
    {(\free{a} + \free{b})\,(1 + \free{a}\,\free{g})\,(1 + \free{b}\,\free{g})\,(1 + \free{a}\,\free{g} + \free{b}\,\free{g})} - 1 \\
 &\qquad \code{ /. }\free{l5} \rightarrow \frac{1}{\free{g}}\Bigg)\code{ // Simplify}
\end{aligned}$}
\end{wlinm}
\begin{wlout}[3]
{{0, 0, 0, 0, 0}, {0, 0, 0, 0, 0}, {0, 0, 0, 0, 0}, {0, 0, 0, 0, 0}, {0, 0, 0, 0, 0}}
\end{wlout}
\noindent which simplifies to $0$ upon taking the parameter \free{q} to its supersymmetric value and relating the extremality parameter \free{mt} to the mass parameter \free{m}. To obtain perfect agreement the gauge freedom $\chi$ is used up. The gauge field goes the same way and comparing fixes the second gauge function, $\lambda$, together with the gauge shift constant $f_0$.

\paragraph{The supersymmetric black hole is a $\text{U}\qty(1)$ fiber over a K\"ahler base} The second kind of interplay is structural rather than a limit. Every supersymmetric solution of five-dimensional minimal gauged supergravity whose Killing-spinor bilinear is timelike is a fibration over a four-dimensional K\"ahler base \cite{Gauntlett:2003fk},
\begin{align}\label{eq:fibration}
  \dd[]{s}_5^2 ={}& - f^2 \qty(\dd[]{y} + \varpi)^2 + f^{-1} \dd[]{s}_B^2 \,,
\end{align}
and the function $f$ that \eqref{eq:f_relation} produced out of thin air is nothing but the inverse of the curvature of that base
\begin{align}\label{eq:f_informative}
  f ={}& - \frac{24}{\ell_5^2 \, R_B} \,,
\end{align}
with $R_B$ its Ricci scalar. The base is a geometry in its own right and deserves its own entry. \sxdata{Curated/Thr__4dE__Kahler-base} is the theory: a four-dimensional Euclidean manifold carrying, besides its metric and frame, an $\mathrm{SU}(2)$ bundle whose sections are a triplet of two-forms $\ensuremath{\mathcal{X}}^{\mathfrak{i}}$, together with a one-form $\ensuremath{\mathcal{P}}$ and the two-form $\ensuremath{\mathcal{R}} = \dd\ensuremath{\mathcal{P}}$. Its \code{\$set} entries are the conditions that make such a manifold K\"ahler, each stored as a named expression that any solution built on the theory can call:
\begin{align}\label{eq:base-conditions}
  \begin{aligned}
  \ensuremath{\mathcal{X}}^{\mathfrak{i}}\ensuremath{\mathcal{X}}^{\mathfrak{j}} ={}& - \delta^{\mathfrak{i}\mathfrak{j}} \mathbbm{1} + \eta^{\mathfrak{i}\mathfrak{j}\mathfrak{l}} \ensuremath{\mathcal{X}}_{\mathfrak{l}} \,, &
  - \frac{1}{2} \ensuremath{\mathcal{X}}^{\mathfrak{i}} \wedge \ensuremath{\mathcal{X}}^{\mathfrak{i}} ={}& \star_{B} \,, &
  \star_{B} \ensuremath{\mathcal{X}}^{\mathfrak{i}} ={}& - \ensuremath{\mathcal{X}}^{\mathfrak{i}} \,, \\
  \dd[]{\ensuremath{\mathcal{X}}^{1}} ={}& 0 \,, &
  \qty(\nabla_{\curly{m}} + \iu \ensuremath{\mathcal{P}}_{\curly{m}}) \qty(\ensuremath{\mathcal{X}}^{2} + \iu \ensuremath{\mathcal{X}}^{3}) ={}& 0 \,, &
  \ensuremath{\mathcal{R}}_{\curly{m}\curly{n}} ={}& \tfrac{1}{2} R_{\curly{m}\curly{n}\curly{p}\curly{q}} \ensuremath{\mathcal{X}}^{1\,\curly{p}\curly{q}} \,,
  \end{aligned}
\end{align}
together with the equation that determines the fibration one-form,
\begin{align}\label{eq:dvarpi}
  \dd[]{\varpi} + \star \dd[]{\varpi} ={}& \frac{\ell_5^3 R_B}{24} \qty(\ensuremath{\mathcal{R}} - \frac{1}{4} R_B \, \ensuremath{\mathcal{X}}^{1}) \,.
\end{align}
The first row of \eqref{eq:base-conditions} says that the $\ensuremath{\mathcal{X}}^{\mathfrak{i}}$ are an anti-self-dual quaternionic triple, normalised against the volume form. The second row is where the gauging shows: only $\ensuremath{\mathcal{X}}^{1}$ is closed, so it alone is a K\"ahler form, $\ensuremath{\mathcal{P}}$ is the connection on the canonical bundle whose curvature $\ensuremath{\mathcal{R}}$ is the Ricci form, and $\ensuremath{\mathcal{X}}^{2} + \iu\ensuremath{\mathcal{X}}^{3}$ is the holomorphic $(2,0)$-form, covariantly constant with respect to $\ensuremath{\mathcal{P}}$.

\sxdata{Curated/Sol__4dE__Kahler-base__Orthotoric} is an explicit geometry that satisfies the above conditions
\begin{align}\label{eq:base-metric}
  \dd[]{s}_B^2 ={}& \ell_5^2 \qty[\frac{x - \rho}{\ensuremath{\mathcal{F}}} \dd[]{\rho}^2 + \frac{\ensuremath{\mathcal{F}}}{x - \rho} \qty(\dd[]{\Phi} + x \dd[]{\Psi})^2 + \frac{x - \rho}{\ensuremath{\mathcal{G}}} \dd[]{x}^2 + \frac{\ensuremath{\mathcal{G}}}{x - \rho} \qty(\dd[]{\Phi} + \rho \dd[]{\Psi})^2] \,,
\end{align}
with the same $\ensuremath{\mathcal{P}}$ and $\varpi$ of \eqref{eq:susy-varpi}. Its notebook computes $\ensuremath{\mathcal{X}}^{\mathfrak{i}}$ from the frame and checks \eqref{eq:base-conditions} and \eqref{eq:dvarpi}. Along the way it stores the base Ricci scalar,
\begin{align}\label{eq:base-ricci}
  R_B ={}& - \frac{\ensuremath{\mathcal{F}}'' + \ensuremath{\mathcal{G}}''}{\ell_5^2 \qty(x - \rho)} \,,
\end{align}
which is \eqref{eq:f_relation} and \eqref{eq:f_informative} in one line.

To see this explicitly we call both solutions and verify that our algebra checks out:
\begin{wlin}
(!\free{sol}!) = GetData["Curated/Sol__5dL__CCLP-susy__Orthotoric"];
Load@(!\free{sol}!)
(!\free{base}!) = GetData["Curated/Sol__4dE__Kahler-base__Orthotoric"];
\end{wlin}
\begin{wlinm}
{\footnotesize
\noindent$\begin{aligned}
 &\free{sol}\code{[\$metric, }\free{gg}\code{, Expression];} \\
 &-\free{f}\code{[]}^{2}\,\big(\code{Diff[}\free{y}\code{[]]} + \fre{\varpi}\code{[]}\big)\otimes
   \big(\code{Diff[}\free{y}\code{[]]} + \fre{\varpi}\code{[]}\big)
   + \free{f}\code{[]}^{-1}\,\free{base}\code{[\$metric, }\free{gg}\code{, Expression];} \\
 &\code{\%\% - \% // Simplify}
\end{aligned}$}
\end{wlinm}
\begin{wlout}[2]
0
\end{wlout}
\noindent The above shows that we have correctly built up the 5d metric.
\begin{wlinm}
\noindent$\begin{aligned}
 &\code{Last@First@}\free{base}\code{[\$metric, }\free{gg}\code{, \$auto, }\free{RicciScalarcd}\code{, Value]} \\
 &\free{f}\code{[]} + \frac{24}{\free{l5}^{2}\;\free{RicciScalarcd}\code{[]}}
   \code{ /. Gen@}\free{sol}\code{[\$function, }\free{f}\code{] /. \%}
\end{aligned}$
\end{wlinm}
\begin{wlout}[1]
0
\end{wlout}
\noindent The first line above is a plain lookup into the fetched association: \code{\$auto} entries are stored under their \code{Value}.
The second line substitutes it into \eqref{eq:f_informative} together with the definition of $f$.

Nothing here was re-computed. Both records were built once, saved once, and the two checks are two substitutions. That is the whole argument of this paper in miniature: a solution stored as code can be \emph{used}, not merely cited.

\paragraph{A more involved interplay of entries: holographic renormalization in 4d} The \code{Curated} corpus also carries a larger set of ``interacting entries'', in which a holographic computation is spread across five of them:
\begin{itemize}\itemsep2pt
  \item \sxdata{Curated/Thr__4dL__Minimal-gauged} --- the bulk theory, four-dimensional minimal gauged supergravity.
  \item \sxdata{Curated/Sol__4dL__Black-hole__Global} --- the dyonic black hole in global coordinates: it solves the equations of motion, then computes its thermodynamic potentials, its electric and magnetic charges, its entropy, and its \emph{divergent} bulk on-shell action.
  \item \sxdata{Curated/Sol__4dL__Black-hole__FG} --- the same solution again, fetched from the previous entry and rewritten in Fefferman--Graham coordinates.
  \item \sxdata{Curated/Thr__3dL__Boundary-minimal-gauged} --- the three-dimensional boundary theory, an entry of its own because the boundary is a different manifold with a different metric.
  \item \sxdata{Curated/Sol__3dL__Induced-black-hole__FG} --- the induced boundary solution, built from the Fefferman--Graham record. Its last section is the holographic renormalisation proper: it adds the Gibbons--Hawking term and the counterterms to the divergent bulk action to obtain the renormalised one, and varies the same combination to obtain the boundary stress tensor and hence the energy.
\end{itemize}
Read these example notebooks in the above order to see the details of the calculation. The final section of \sxdata{Curated/Sol__4dL__Black-hole__Global} fetches the boundary record, imports the renormalised action and the energy from it, and verifies the quantum statistical relation.

\paragraph{Browsing the available entries} \sx{ShowData}\code{[]} is a simple interactive browser, allowing a bird's-eye view of the available entries:
\begin{wlin}
ShowData[]
\end{wlin}
\begin{wloutm}
\vspace{0.7ex}\centering
\begin{sxdpanel}
\noindent\sxdtabplain[sxdbrown]{\sxdpencil\ Vasko}\;\textcolor{sxdofftext}{\code{\small|}}\;%
\sxdtab{Curated}\;\textcolor{sxdofftext}{\code{\small|}}\;%
\sxdtabplain{Vasko-bot}\hspace{1.5em}\sxdfield
\par\vspace{0.9ex}
\setlength{\tabcolsep}{1.5pt}\renewcommand{\arraystretch}{1.4}
\noindent\begin{tabular}{@{}l@{\hspace{0.3em}}l l l l l@{}}
\sxdname{Curated/Sol__3dL__Induced-black-hole__FG} & \sxdbtn{OpenData} & \sxdbtn{CopyName} & \sxdbtn{CopyData} & \sxdbtnoff{DeleteData} & \sxdbtnoff{CurateData} \\
\sxdname{Curated/Sol__4dE__Kahler-base__Orthotoric} & \sxdbtn{OpenData} & \sxdbtn{CopyName} & \sxdbtn{CopyData} & \sxdbtnoff{DeleteData} & \sxdbtnoff{CurateData} \\
\sxdname{Curated/Sol__4dL__Black-hole__FG} & \sxdbtn{OpenData} & \sxdbtn{CopyName} & \sxdbtn{CopyData} & \sxdbtnoff{DeleteData} & \sxdbtnoff{CurateData} \\
\sxdname{Curated/Sol__4dL__Black-hole__Global} & \sxdbtn{OpenData} & \sxdbtn{CopyName} & \sxdbtn{CopyData} & \sxdbtnoff{DeleteData} & \sxdbtnoff{CurateData} \\
\sxdname{Curated/Sol__5dL__CCLP__Boyer-Lindquist} & \sxdbtn{OpenData} & \sxdbtn{CopyName} & \sxdbtn{CopyData} & \sxdbtnoff{DeleteData} & \sxdbtnoff{CurateData} \\
\sxdname{Curated/Sol__5dL__CCLP-susy__Orthotoric} & \sxdbtn{OpenData} & \sxdbtn{CopyName} & \sxdbtn{CopyData} & \sxdbtnoff{DeleteData} & \sxdbtnoff{CurateData} \\
\sxdname{Curated/Thr__3dL__Boundary-minimal-gauged} & \sxdbtn{OpenData} & \sxdbtn{CopyName} & \sxdbtn{CopyData} & \sxdbtnoff{DeleteData} & \sxdbtnoff{CurateData} \\
\sxdname{Curated/Thr__4dE__Kahler-base} & \sxdbtn{OpenData} & \sxdbtn{CopyName} & \sxdbtn{CopyData} & \sxdbtnoff{DeleteData} & \sxdbtnoff{CurateData} \\
\sxdname{Curated/Thr__4dL__Minimal-gauged} & \sxdbtn{OpenData} & \sxdbtn{CopyName} & \sxdbtn{CopyData} & \sxdbtnoff{DeleteData} & \sxdbtnoff{CurateData} \\
\sxdname{Curated/Thr__5dL__Minimal-gauged} & \sxdbtn{OpenData} & \sxdbtn{CopyName} & \sxdbtn{CopyData} & \sxdbtnoff{DeleteData} & \sxdbtnoff{CurateData} \\
\sxdname{Curated/Thr__5dL__STU-gauged} & \sxdbtn{OpenData} & \sxdbtn{CopyName} & \sxdbtn{CopyData} & \sxdbtnoff{DeleteData} & \sxdbtnoff{CurateData} \\
\end{tabular}
\end{sxdpanel}
\end{wloutm}
\noindent The three names at the top of the panel are \emph{aliases}\footnote{Next to the alias tabs sits the search field. The query is split into words, and an entry stays listed when every word occurs, case-insensitively, in its \code{alias/name}. The double-underscore names make this plain substring matching effectively structural: \code{5dL CCLP} finds exactly the CCLP records.}. An alias is a folder of entries with somebody's name on it. Each user chooses their alias once, when they first run \sx{Welcome}\code{[]}, and from then on everything they save lands under it. Reading is unrestricted: any alias present in the data directory can be fetched. Writing, by design, is restricted: one writes only under one's own alias. In the example panel above \code{Vasko} is the current user, signified by the pencil; \code{Curated} is a special alias where all the curated entries (think of them as ``official'') are stored; and \code{Vasko-bot} is a second alias, belonging to the same author, where \code{Vasko}'s agents store their entries.
Depending on their permissions, a user can \sx{OpenData} (open the corresponding notebook), \code{CopyName} (copy the name of the entry), \sx{CopyData} (copy someone else's entry under their own alias), \sx{DeleteData} (delete a data entry) and \sx{CurateData} (copy one of their own entries into the curated corpus).

\section{Benchmarking: \solx vs \xcoba}
\label{sec:bench}

Any new package owes the reader a comparison with what already exists. For \sx{Compute} the nearest thing to compare against is \xcoba's \href{http://www.xact.es}{\code{MetricCompute}}, which builds the tensor components of a chart out of a component metric. This section reports a head-to-head between the two on two five-dimensional records, one of them the solution of \cref{sec:life}\footnote{The benchmark harness is not part of the package; it is available upon request.}.

\subsection{Flexibility}
\code{MetricCompute} produces a fixed set of objects: the metric in both index placements and its determinant, the Christoffel symbols all-lower and once-raised, the Riemann tensor all-lower and once-raised, the Ricci tensor and the Ricci scalar. Which of them are actually built is decided by a dependency table inside \xcoba, and the first and second derivatives of the metric are built on the way whether or not anybody wanted them. Simplification is a single function, handed over through the \code{CVSimplify} option and applied to every component, with a boolean deciding whether it runs in parallel.

\sx{Compute} has no such menu. It computes whatever the record's \code{Chain} names, in whichever bundle the indices live, along whichever route the author wrote down, and it passes every component through the two sieves of \sxD{Apply}, one serial and one parallel. \cref{sec:compute} used all of that: the vielbein carries a chart index and a frame index at once; the spin connection is written straight from its definition; the $\gamma$-matrices are computed in two bases; and the serial sieve carries a substitution that the distributed one cannot express.

This says nothing about what \xact can or cannot do, in principle. \xcoba exposes the primitives, so every array \sx{Compute} produces can be re-created by hand-written code using vanilla \xact. Our contribution is devising a unified syntax to achieve that with ease.

\subsection{Need for speed \textit{and} quality}
Both engines make clever use of tensor symmetries and allow the user to run calculations in parallel. In order to compare them, we focus on where they overlap: namely, on computing the metric-associated tensors. All timings in this paper come from the same laptop, an Apple M1 Pro with 8 cores and 32\,GB of RAM, with nothing else running on it; the runs of this section use Wolfram 15.0.1 with eight parallel subkernels.

To ensure a fair benchmark both engines are given the same input, allowed the same means, and asked for the same output. 
The input is the array of the metric components with both indices down: $\code{gg}_{\ichart{\curly{m}}\ichart{\curly{n}}}$. The output is nine arrays:
\begin{align*}
\code{gg}_{\ichart{\curly{m}}\ichart{\curly{n}}},\;\;
\code{gg}^{\ichart{\curly{m}}\ichart{\curly{n}}},\;\;
\wldensityupp{\code{g}}\code{g},\;\;
\Gamma[\nabla]_{\ichart{\curly{m}}\ichart{\curly{n}}\ichart{\curly{o}}},\;\;
\Gamma[\nabla]\indices{^{\ichart{\curly{m}}}_{\ichart{\curly{n}}\ichart{\curly{o}}}},\;\;
\code{R}[\nabla]_{\ichart{\curly{m}}\ichart{\curly{n}}\ichart{\curly{o}}\ichart{\curly{p}}},\;\;
\code{R}[\nabla]\indices{_{\ichart{\curly{m}}\ichart{\curly{n}}\ichart{\curly{o}}}^{\ichart{\curly{p}}}},\;\;
\code{R}[\nabla]_{\ichart{\curly{m}}\ichart{\curly{n}}},\;\;
\code{R}[\nabla]\,.
\end{align*} Both simplify with \code{Simplify[Together[\#]] \&}, applied once per independent component, in parallel, with no serial pre-pass on either side. Both declare the same constants and functions to \xact beforehand, which matters because a declaration is what lets \wlref{Simplify} treat a symbol as a quantity an assumption can be about. Importantly, both engines agree on every tensor component, up to \wlref{Simplify}.
Agreement in value is not by itself enough, because an engine can be fast by simplifying less. The control is \wlref{LeafCount}, which counts the atoms of an expression, so the same quantity simplified further has fewer of them. Across all runs the summed leaf counts agree to within $0.015\%$, practically the same. 
One thing the two engines do not share is the environment their simplifier runs in. \solx ships the record's assumptions into the subkernels, together with the definitions of the declared symbols. \xcoba's parallel simplification is hard-wired to ship neither, and its interface offers no way to ask for it, so the assumption-bearing runs of \code{MetricCompute} were produced by supplying both from \emph{outside the package}. In a way we have given \code{MetricCompute} a custom-designed ``crutch'' to allow it to work with \code{\$Assumptions}.

The comparison is presented in \cref{tab:bench}, on \sxdataC{Sol__5dL__CCLP__Boyer-Lindquist} and \sxdataC{Sol__5dL__CCLP-susy__Orthotoric}. Besides the engine, the table varies two things. One is the assumptions. \solx always ships them to the simplifier and \xcoba never does, so neither setting is neutral ground, and both are measured for both engines. The other is the route. In the middle block \sx{Compute} is driven the way \xcoba drives itself, with the derivatives of the metric built as explicit arrays and every tensor assembled out of them, the whole recipe written as an ordinary \code{Chain}. That block separates the cost of the route from the cost of the engine.

\begin{table}[h]
\centering
\begin{tabular}{@{}lrrrr@{}}
\toprule
 & \multicolumn{2}{c}{\small\sxdataC{Sol__5dL__CCLP__Boyer-Lindquist}} & \multicolumn{2}{c}{\small\sxdataC{Sol__5dL__CCLP-susy__Orthotoric}} \\
\cmidrule(lr){2-3}\cmidrule(lr){4-5}
 & no assumptions & assumptions & no assumptions & assumptions \\
\midrule
\multicolumn{5}{@{}l}{\code{Compute}, the record's own chains} \\
\quad wall clock (s) & \losenum{65.81} & \losenum{92.87} & \losenum{982.69} & \winnum{2398.16} \\
\quad leaf count & \leafnum{727\,000} & \leafnum{727\,000}
                 & \leafnum{2\,662\,078} & \leafnum{2\,662\,078} \\
\addlinespace
\multicolumn{5}{@{}l}{\code{Compute}, mimicking \xcoba} \\
\quad wall clock (s) & 87.32 & 114.51 & 1013.65 & 2386.76 \\
\quad leaf count & \leafnum{727\,012} & \leafnum{727\,012}
                 & \leafnum{2\,662\,080} & \leafnum{2\,662\,080} \\
\addlinespace
\multicolumn{5}{@{}l}{\code{MetricCompute}} \\
\quad wall clock (s) & \winnum{41.28} & \winnum{85.78}$^\dagger$ & \winnum{635.87} & \losenum{2905.69}$^\dagger$ \\
\quad leaf count & \leafnum{726\,891} & \leafnum{727\,028}
                 & \leafnum{2\,662\,150} & \leafnum{2\,662\,138} \\
\bottomrule
\end{tabular}
\caption{Nine tensor arrays, computed from the same metric with the same
simplification function, on \sxdataC{Sol__5dL__CCLP__Boyer-Lindquist}
and \sxdataC{Sol__5dL__CCLP-susy__Orthotoric}. Wall clock in seconds,
and under it the summed \wlref{LeafCount} of all nine arrays. Every run
reproduces the stored record exactly. In each column the faster of
\code{Compute} on the record's own chains and \code{MetricCompute} is set in
green, the slower in red. The mimicking block is reported as an aside and does not participate in the race.
The $\dagger$ signifies that \code{MetricCompute} has no
way of sending assumptions to its subkernels, so those two runs were given
them from outside the package.}
\label{tab:bench}
\end{table}
\noindent Without assumptions \sx{Compute} loses to \code{MetricCompute} by a factor of 1.59 on \sxdataC{Sol__5dL__CCLP__Boyer-Lindquist} and by 1.55 on \sxdataC{Sol__5dL__CCLP-susy__Orthotoric}. With assumptions the picture turns: \sx{Compute} is practically level on the first record and ahead by 1.21 on the second. What the engines do differently, underneath, is choose which components the simplifier sees. \sx{Compute} hands over only the components that the record's symmetries leave independent. \code{MetricCompute} hands over nearly twice as many\footnote{977 against \sx{Compute}'s 502, over the nine arrays. The excess is components that its own symmetries already fix, plus the derivatives of the metric that its route builds on the way.}, but the extra ones are the easy ones, so its average component is cheaper. Each engine thus holds one advantage: \sx{Compute} simplifies fewer components, \code{MetricCompute} pays less per component. Without assumptions the per-component advantage is the larger of the two, and \code{MetricCompute} wins. Shipping the assumptions makes every component dearer on both sides by a comparable amount, which shrinks the per-component advantage; the shorter list then pulls \sx{Compute} level on the first record and ahead on the second. The middle block makes two points. \xcoba's particular recipe is one concrete \code{Chain} one can choose for \sx{Compute}. Driving that \code{Chain}, instead of the ``native'' one, moves the clocks only modestly. Thus, the route is not what separates the engines; the simplification is. 

We want to disclose that for the particular experiments in \cref{tab:bench} carrying the assumptions is ``dead weight'': the final \wlref{LeafCount} is practically the same with or without assumptions. That is because in the present examples the metrics see no radicals. Using assumptions really pays off only when radicals are present. In the \sxdataC{Sol__5dL__CCLP-susy__Orthotoric} example that is: the frame, the spin connection, and the $\gamma$-matrices. Said differently: if there were a \code{MetricCompute} analogue for
those tensors it would not simplify nearly as well as \sx{Compute}. The tests above are a proxy for this behaviour, based on what \xcoba currently provides. Note that the desire to properly distribute assumptions across kernels is not cosmetic. If we specified no assumptions whatsoever we would finish computing the tensor components faster, but then, when we pass those answers to the Killing spinor equation of \cref{sec:verify},
\mma would struggle to make a dent in \wlref{Simplify}-ing it for a very long time, rendering the results not practically useful.

In summary: on the nine metric-related tensors without assumptions \code{MetricCompute} is faster. In the regime where \solx actually runs --- with assumptions on, \sx{Compute} comes out ahead. The real benefit of that is visible beyond the metric sector. Both engines leave room for improvements. To spell out a few possible directions:
\begin{itemize}
  \item \code{MetricCompute} could handle assumptions better: the setting that keeps \code{\$Assumptions} and the declared symbols out of its subkernels is currently not configurable.
  \item Both engines could make explicit use of the isometry group of the ansatz; neither consults it today.
  \item \sx{Compute} could let the user decline the assumptions where they are dead weight; at present they are always shipped.
  \item \code{MetricCompute} could simplify fewer components: part of its run goes to components its own symmetry declarations already fix, and to intermediates the user never asked to see.
  \item Both engines parallelise one simplifier call per component, so a run ends waiting on its few largest components; splitting this across subkernels is an open problem.
\end{itemize}

\section{\solx and agents}
\label{sec:agents}

A record that replays on a fresh kernel, a corpus of finished examples, and a documentation whose every example is executed were all designed with a human user in mind. That turns out to be what an AI agent needs even more urgently than a human does: a template that shows what finished work looks like, and a kernel that says whether a claim is true before the claim is made. This section reports what happened when we asked an agent to use \solx by itself. We wrote a small toolkit that clears the obstacles standing between an agent and the package, described in \cref{sec:kit}, and then gave an agent three tasks to work through with it, reported in \cref{sec:runs}. We close with what we expect agents to be able to do in the future. Everything below assumes route~B of the installation instructions of \cref{sec:installation}: the repository, not the paclet alone. The repository carries the notebooks that the agents ``learned from'', together with the notebooks they produced. The directory \code{agent/} is the toolkit of this section, self-documented in its own \code{README} and maintained with the package.

A fair question before any toolkit discussion: why not paste the problem into a chat with a state-of-the-art model and ask it to check the algebra carefully, step by step? Because what comes back is a transcript: plausible algebra that has to be re-derived to be trusted, which is the artifact this paper set out to replace. Scale compounds the problem: verifying the solutions of \cref{sec:bench} means handling expressions whose combined size runs to millions of leaves, arithmetic that no model performs reliably. In contrast, the kernel performs exact verifiable algebraic manipulations. The workflow below therefore splits the work: the model chooses what to try, the kernel decides what is true, and what the agent hands back is a replayable entry rather than prose. We provide no controlled chat-versus-\solx measurement. 
Instead, we report that without the toolkit an agent tasked with what became runs 1 and 2 ran some twelve hours and did not reach the goal. With the toolkit the same scope was completed, correctly, in under three hours.

The toolkit assumes a shell (macOS or Linux, \code{bash}) and an agent that can run shell commands and read files. An agent confined to a notebook evaluator is not supported, for a reason worth spelling out. When an \xact session goes wrong (a symbol defined twice, packages loaded in the wrong order) the accumulated definitions cannot be reliably undone from within the session; the one dependable recovery is to quit the kernel and start over. From the shell that is a two-second command. An agent that lives inside a single kernel has no way to issue it.

\subsection{The agent toolkit}
\label{sec:kit}
The toolkit's mission fits in one sentence: read the existing corpus for methods and for how your user writes, read a paper, and author entries your user could have written. Three obstacles stand between an agent and that mission. First, a full \xact stack load costs the better part of a minute, so the kernel must survive between calls. Second, the corpus is notebooks and the deliverable is a notebook, while an agent reads and writes plain text. Third, a human squints at a suspicious output and re-checks; an agent records it as truth. So failures have to be made loud, and claims have to come with the evidence to audit them. Each item below addresses one of the three.
\begin{itemize}
  \item \code{sxk}\footnote{The name abbreviates \emph{SolutionsX kernel}.} drives one long-lived kernel from the shell, and it simulates ``using notebook cells'': each submission is one cell, evaluated in that kernel, with the output returned as text. \code{sxk try} runs a submission under a time budget; on expiry the submission is cut off and the kernel, with all its state, survives. \code{sxk again} is the recovery verb: it restarts the kernel and re-runs a startup file the agent maintains (load the package, set the identity, reload the entry), so a session that has gone wrong is abandoned without losing its footing. The channel runs both ways: the user can interrupt the agent's loop at any moment by typing \code{sxk tell "<message>"} in a terminal, and every subsequent \code{sxk} command shows the message to the agent until the agent acknowledges it.
  \item \code{nb2txt} and \code{nb-author} exist because the corpus is notebooks and the deliverable is a notebook, while an agent reads and writes text. \code{nb2txt} extracts from a notebook exactly what its human author typed (input and text cells), and nothing the kernel generated (output cells). \code{nb-author} is the way back: from the agent's typed session it builds a notebook of input cells only, which the user then runs top to bottom themselves, watching every output appear on screen.
  \item \code{reporter} exists because a result produced in the long-lived \code{sxk} kernel may silently depend on everything that kernel evaluated before it. Its job, in plain terms: run a given script on a brand-new kernel that has seen nothing else, and hand back the full transcript, with every line that looks like an error or warning pulled out. It renders no verdicts; judging the transcript is the agent's job. The idea behind this is that: ``it worked in my temporary kernel'' is a weak statement that is easily corrupted. In contrast: ``it works on a fresh kernel, here is the transcript attached'' is a claim the user can audit.
  \item \sx{WorkAs} is the identity verb, and the one piece that lives in the package rather than the toolkit. The first thing an agent runs is \code{WorkAs["Vasko-bot"]}: it switches to the agent's own alias, drops the curator flag, leaves the user's configuration file untouched, and verifies that the switch took. That is to ensure that the agent does not accidentally write in under the user alias.
  \item \sx{Compute} gains a progress channel: it writes one line per finished component into a log file, and \code{sxk} displays the latest line together with how long ago it was written. This is the shell's substitute for the progress monitor a notebook shows, and it doubles as the stall detector: a kernel that is busy while its latest line keeps getting older is stuck on one component. This gets flagged and the agent is forced to reconsider the computation approach.
  \item \code{METHODS.md} and \code{TRAPS.md} are hint files, keyed by symptoms the agents experienced during real runs --- ``the hard-won lessons''. \code{METHODS.md} holds approach hints, each with the cost record that earned it. \code{TRAPS.md} holds the silent failures: the cases where the kernel or the package returns a wrong or empty answer with no error. Together they are a rudimentary continuous-learning protocol: a setback in one run is written down as a hint, and every later run reads the hints before hitting the same wall. \Cref{sec:runs} contains one full turn of that cycle.
  \item \code{LADDER.md} is policy. It defines what counts as a structurally different attempt on an obstacle, what counts as a verified fact, and how many failed attempts and how much kernel time one obstacle is worth. Past those thresholds the agent must stop working on the obstacle and write a report for the user, whose most important line is the single question whose answer would unblock it.
\end{itemize}
The working contract, in one line: cheap attempts, loud failures, auditable claims.

The budgets and thresholds are configuration, not doctrine, and the user controls them. They live in one flat text file, \code{agent/kit.conf}, which the user edits like any other file; a single value can also be overridden for one session from the terminal, with \code{sxk set}, and \code{sxk config} prints every effective value and where it came from. The full list is in \cref{app:kit}. The budget levers are enforced by \code{sxk} itself; the stall thresholds are advisory, because nothing can force an agent to comply. What the toolkit cannot enforce, it makes visible.

\subsection{Example blind runs}
\label{sec:runs}
With the toolkit written down, we gave an agent three tasks. Each run saw the cloned repository, the toolkit, and its task, and none of our earlier attempts\footnote{The runs' journals, reports and resource ledgers are available on request, like the harness of \cref{sec:bench}.}. Run 0 exercises the from-scratch path; runs 1 and 2 take on a published paper, the accelerating and supersymmetric AdS$_4$ black holes of \cite{Cassani:2021dwa}, starting from its arXiv source. \Cref{tab:tasks} lists the three tasks, as the runs' journals record them, and under each task the entries the run produced: seven in total, five authored and two theories recycled from the human's corpus with \sx{CopyData}.

\begin{table}[h]
\centering
\small
\begin{tabular}{@{}p{0.95\textwidth}@{}}
\toprule
run 0: \textit{Build the theory of three-dimensional Einstein gravity with a negative cosmological constant from nothing, and show that the BTZ black hole solves its equations of motion.} \\[2pt]
\quad \sxdata{Vasko-bot/Thr__3dL__Einstein-AdS} \\
\quad \sxdata{Vasko-bot/Sol__3dL__BTZ__Global} \\
\midrule
run 1: \textit{Recreate sections 2.1 and 2.2 of \cite{Cassani:2021dwa}, the accelerating black hole of minimal gauged supergravity and the regularity analysis of its spindle horizon, starting from the paper's arXiv source.} \\[2pt]
\quad \sxdata{Vasko-bot/Sol__4dL__Accelerating-black-hole__PD} \\
\quad the theory \sxdatatext{Thr__4dL__Minimal-gauged}, recycled from the human's corpus \\
\midrule
run 2: \textit{Recreate the same paper's section 3, the holographic thermodynamics, per a written brief: the solution induced on the cutoff slice, the conformal boundary solution, and the thermodynamic sections appended to run 1's entry.} \\[2pt]
\quad \sxdata{Vasko-bot/Sol__3dL__Induced-accelerating-black-hole__ADM} \\
\quad \sxdata{Vasko-bot/Sol__3dL__Boundary-accelerating-black-hole__Conformal} \\
\quad four sections appended to run 1's entry \\
\quad the theory \sxdatatext{Thr__3dL__Boundary-minimal-gauged}, recycled from the human's corpus \\
\bottomrule
\end{tabular}
\caption{The three tasks, as the runs' journals record them, lightly
trimmed, and under each task the entries the run produced under the
agent's alias. Recycled means copied unmodified from the human's corpus
with \sx{CopyData}.}
\label{tab:tasks}
\end{table}
Every authored entry was replayed on a fresh, isolated kernel, and the paper-facing ones were checked against the published equations one by one, with the AdS radius kept general where the paper sets $\ell = 1$. Run 2's record is the deepest: the boundary metric and gauge field at the cutoff, the cancellation of both divergent orders of the counterterm bracket in an ADM slicing with shift (no template for this existed in the corpus), a traceless boundary stress tensor with its Ward identities, all four charges, the renormalised on-shell action, the quantum statistical relation, the Smarr formula, and the first law with all eight residuals vanishing identically. 

\cref{tab:agents} is the campaign's ledger. Its sharpest reading is that the kernel time is the cheap part; the expensive part is deciding what to ask the kernel. Productive tensor algebra took twelve minutes against three hours and sixteen minutes of wall clock, and starting kernels cost more than every Ricci tensor, variation and simplification combined; for most of the clock no kernel was busy at all, the agent being occupied with reading the corpus, choosing an approach, authoring notebooks and verifying. The table also shows the kernel spending as much time on attempts that timed out as on attempts that worked, and that is expenditure, not waste: a budgeted attempt costs its budget, leaves the kernel alive, and returns a verdict, and those verdicts are what located each obstacle precisely enough for the method that dissolved it to be found. 
\begin{table}[h]
\centering
\begin{tabular}{@{}lrrr@{}}
\toprule
 & run 0 & run 1 & run 2 \\
\midrule
wall clock & ${\approx}\,35$ min & $40$ min & $2$ h $01$ min \\
productive kernel time & ${\approx}\,1$ min & ${\approx}\,2$ min & $9$ min $10$ s \\
kernel time in budgeted timeouts & ${\approx}\,0$ & ${\approx}\,5$ min $30$ s & $9$ min $10$ s \\
full \xact-stack kernel loads & 9 & 3 & 14 \\
\code{sxk} submissions & 57 & 95 & 393 \\
front-end utility launches & 12 & 7 & 17 \\
context-counter tokens & ${\approx}\,310$k & 80--100k & ${\approx}\,308$k \\
entries authored (recycled) & 2 & 1 (+1) & 2 (+1) \\
setbacks, all recovered & 4 & 5 & 6 \\
\bottomrule
\end{tabular}
\caption{Time and resources of the three runs, transcribed from their
records: times from file modification times and in-kernel timings, counts
from the session logs. The token counter is the agent harness's context
counter, read by the agent at the time: its exact semantics are opaque to
us and it is not billed API volume, so it is useful for the relative
weight of the runs, not for cost; run 1's figure is an estimate
cross-checked against the other two. Run 0's wall clock spans its two
turns and excludes a ten-minute orientation turn in which the agent read
the toolkit's documentation.}
\label{tab:agents}
\end{table}

Where the runs struggled is more instructive than where they cruised. Run 0 produced the one failure that no guard caught. The agent had copied a cell out of a corpus notebook, and the front end had long ago hard-wrapped that cell's longest product at the width of the screen. Read back as plain text, the inserted line break split the product into two syntactically complete expressions, the variational derivative acted on the constant first half, returned zero, and a zero equation of motion was stored with no message anywhere. It surfaced minutes later, as a solver complaining about a degenerate system. The repair took four minutes, and the incident is now both a \code{TRAPS.md} entry and a warning wired into the authoring tool: one run's setback, become the next run's hint.

In run 1, Einstein's equations verified in sixteen seconds and Maxwell's would not. Three structurally different attempts were made, each hit the 110-second budget. The first exit is the one the ladder prescribes at the threshold: stop grinding and ask what the computation is a check \emph{of}. For this ansatz Maxwell's equations hold with the structure functions kept \emph{arbitrary}: they are satisfied by the shape of the metric and gauge field alone, before any particular solution is chosen. Substituting the explicit quartic structure functions had therefore been gratuitous work; kept symbolic, the check ran in seconds and proves that stronger, ansatz-level statement. The one human intervention during the campaign attacked the same obstacle. It was a prompt suggesting to
remove the dependence on trigonometric functions by exchanging the polar coordinate $\theta$ for $x$ as
$\cos\theta = -x$, which makes every metric function rational. Rebuilt in that chart, the obstacle disappeared: the check that had timed out three times ran in a tenth of a second\footnote{Every slow step sped up with it, by roughly three orders of magnitude. This was the largest single speed-up of the campaign, and it came from the human, not from the agent's next attempt. It is why \code{sxk} keeps the two-way channel open at all times.}.

Run 2's hardest obstacle was solved by the agent alone. The boundary stress tensor is assembled out of ingredients that are each small, a couple of hundred leaves apiece. Combining them, with the exact data of the cutoff slice substituted, produces one rational expression in seven parameters that is too large to bring to a common denominator in any reasonable time: three structurally different attempts at simplifying it hit their budgets and were cut off. The method that worked never forms that expression. First expand formally: write each function on the cutoff slice as a series in the cutoff with undetermined coefficients, so that each order of the counterterm bracket becomes a compact formula in those coefficients, and the divergent orders cancel identically at the level of the formulas. Then substitute: compute one inexpensive series per slice function and feed its coefficients into the per-order formulas. Each step fit comfortably inside the budget that the direct route had blown three times.

The human's hand in the campaign is three prompts and two interruptions, and nothing else:
\begin{itemize}\itemsep2pt
  \item the three prompts of \cref{tab:tasks};
  \item between run 0's two turns: an input-formatting fix, and a request to extend the entry;
  \item during run 1: the mid-run interruption hinting to change coordinates.
\end{itemize}
The stall ladder never fired; every obstacle was overcome before the thresholds compelled a report. Silent failures did occur, seven of them, wrong or empty answers with no error. Six were caught by discipline the hint files had already written down; the seventh, the stored zero above, is why it is written down now. For calibration: an earlier attempt at the same paper, made before the toolkit and its hint files existed, ran some twelve hours and was abandoned with nothing meaningful to show. The same paper, the same package, the same kind of agent; what changed in between is the toolkit.

None of this is a mature workflow. The toolkit is some seven hundred lines of shell and three text files; the corpus the agent read holds under twenty records. What we expect is that two effects will compound to significantly better results and a better user/agent experience: every codified paper enlarges the set of worked examples the next agent reads, and every setback sharpens the hint files, so later runs should need less time and less supervision than the earlier ones. The direction this points in is: agents that contribute rather than merely ``codify papers''. \solx plays a crucial part in that as it is the verifier: an agent's ideas, like anybody's, are mostly wrong, and \solx is the machine that tells it so before it tells you.

\section{Discussion}
\label{sec:conclusions}
In this paper we have introduced \solx, a new \mma package based on \xact. The main motivation behind creating this package is the desire to turn existing papers and novel calculations into verifiable, re-usable code. \solx provides two main contributions on top of \xact: a unified database of \xact environments, together with a workflow to call, load and operate with them; and a flexible way to compute the components of tensors living on multiple vector bundles simultaneously.
The package automatically ports \xact's abilities to work with:
\begin{itemize}
  \item abstract tensors (\xtensor): tensor symmetries, generalized connections/covariant derivatives, tensor rules, tensor densities, Lie derivatives, variational derivatives
  \item explicit coordinate charts (\xcoba): coordinate transformations of tensors
  \item differential forms (\xterior): forms, exterior derivative,  Cartan structure equations, Lie derivatives, inner contractions, variational derivatives of forms
  \item spinors in arbitrary dimensions (\fieldsx): inner bundles with metric, spin structure, spin connection, gamma matrices, spinors, Fiertz identities, BRST operators, cohomology
\end{itemize}
All of this functionality allows one to work with \solx on a large class of geometry/supergravity/QFT problems. The examples which ship with the package focus on explicit supergravity solutions, but that is just a sliver of what one can achieve with \xact. We would like to categorize the class of problems amenable to \solx-codification in three categories: (\cls{A}{A}) \textit{working today}, (\cls{B}{B}) \textit{can be made to work with minimal extension to \xact}, (\cls{C}{C}) \textit{need major extension to \xact}.

\begin{classtable}{A}{working today}
  Working with Lagrangian theories & Equations of motion, checking closure under supersymmetry variations, Killing spinor equations. Within \solx such problems will fall into the \code{Thr} category as no explicit chart will be defined \\
Working with higher derivative Lagrangian theories & This should really be bundled together with the previous class of problems as it requires no new functionality, but we want to explicitly emphasize it, especially because the algebra gets progressively more involved, which is where a CAS starts to shine \\
Working with explicit solutions of supergravity or devising fixed curved backgrounds on which one can place a QFT/SQFT/SCFT & Within \solx such problems are usually extending a \code{Thr} entry with a local chart, thus making it a \code{Sol} \\
Holographic renormalization: Fefferman--Graham expansions and induced boundary data & Within \solx this is a chain of related entries: the bulk \code{Sol}, its Fefferman--Graham chart as a second \code{Sol}, and the induced boundary solution as a third. The boundary stress tensor of the agent runs in \cref{sec:runs} was assembled this way \\
\end{classtable}

\begin{classtable}{B}{minimal extension to \xact}
 Supersymmetric localization \`a la Pestun \cite{Pestun:2007rz} & In such a problem \solx can already handle: the rigid background, the off-shell closure of the algebra, the $\mathcal{Q}V$ deformation. What it cannot yet do is one-loop determinants and instantons, see class \cls{C}{C}. \\
 Equivariant localization in supergravity \cite{BenettiGenolini:2023kxp,Martelli:2023oqk} & The needed technical extension is to introduce polyforms and the equivariant exterior derivative of the Cartan model $\dd_X = \dd - \iota_X$, nilpotent on $X$-invariant forms since $\dd_X^2 = -\mathcal{L}_X$ \\
 Dimensional reductions \& consistent truncations in supergravity & Take for example reducing 11d supergravity on $S^4$ \cite{Nastase:1999cb}. This can be achieved by a simple interplay between three \solx entries: one for the 11d space, one for the $S^4$ and one for the 7d space together with its vectors and scalars. The required functionality to make this actually pleasant to codify is a rudimentary list of rules that automatically generates the field content of the 7d theory, see class \cls{C}{C} \\
 Verifying explicit instanton gauge fields obtained from the ADHM construction \cite{Atiyah:1978ri} & Solving the ADHM constraints is finite matrix algebra and needs only \mma itself. The resulting gauge field on an $\mathbb{R}^4$ chart is then a \code{Sol} like any other: self-duality, the Yang--Mills equations and the topological charge density are component checks. On paper this needs no extension at all; it sits in class \cls{B}{B} because a nonabelian gauge field through a \fieldsx inner bundle has not been exercised in \solx. The general-$k$ proof of self-duality and the integration over the instanton moduli space are not claimed; for the latter see class \cls{C}{C} \\
\end{classtable}

\begin{classtable}{C}{major extension to \xact}
  Dimensional reductions \& consistent truncations - done properly & That would require teaching \xact group theory and Lie algebras. A possible route will be integration with \code{LieART} \cite{Feger:2019tvk}. \\
 Exceptional geometry & Exceptional field theory organizes the supergravity fields into representations of $E_{d(d)}$, with a generalized Lie derivative in place of the ordinary one \cite{Berman:2020tqn}. The calculus is finite but it is representation theory throughout, so this entry stands or falls with the group-theory layer of the previous one. \\
 Supersymmetric localization - done properly & On a general rigid background the class \cls{B}{B} algebra ends at the BPS locus; what follows is an equivariant index engine, in two programmable halves. The geometric half is tensor algebra: the zeros of the supercharge's Killing vector, and the linearized field content around the locus (\code{xPert} territory). The index half needs no \xact: the weights of the isometry action at each fixed point are finite linear-algebra data, and the Atiyah--Bott formula turns them into the index and hence the one-loop determinant \cite{Pestun:2007rz,Pestun:2016zxk}; instanton blocks glue over the same fixed points \cite{Nekrasov:2002qd}. A long shot, but every step is finite, and the input is what an entry already stores. \\
 Spectra and perturbations & \xact already covers the algebraic part: \code{xPert} builds metric perturbation theory to any order \cite{Brizuela:2008ra}. The reduction to master equations \cite{Jansen:2019wag} and the spectral step of solving them are the missing layers; for the latter numerical tools exist outside \xact \cite{Jansen:2017oag}. \\
\end{classtable}
\noindent The tables above make it clear that \xact + \solx are already powerful tools in the quest of codifying calculations on a large array of problems. They also make it clear that there is a long list of homework in extending the tools to capture a larger set of problems and research programs\footnote{The extension suggestions are limited by the author's own understanding; feel free to email me yours.}.

We would like to use this paper as an opportunity to call out to ``programmatically inclined'' physicists to help develop some of these extensions: if that is in your ballpark, email me at \solxmail.

The second message of this paper is that once problems are codified in a verifiable fashion, agents suddenly become far more useful compared to ``ordinary chat''. We experimented with giving \href{https://claude.com/claude-code}{\code{Claude Code}} the task of recreating a paper \cite{Cassani:2021dwa} and it managed to codify it correctly in under three hours, see \cref{sec:runs}. In order to aid its work we developed a simple agent toolkit, which ships with the GitHub version of \solx. The promise here is worth emphasising. An agent who is given access to a good database and an independent verifier performs well on re-creating existing literature. More importantly, this might be a pathway to allow agents to operate on complex open problems in hep-th by separating the task in two: the thinking, ideating part (agents are already pretty good at that), and the continuous testing, fidgeting and verifying of their claims with standard code. \solx provides the framework for the second part 
when it comes to supergravity/QFT related problems. Still, the agent toolkit that we have provided is in a somewhat rudimentary, experimental stage.

We would like to use this paper as an opportunity to call out to ``physically inclined'' developers to help streamline the database, make \solx smoother to use and develop a more sophisticated agent toolkit: if that is in your ballpark, email me at \solxmail.

Finally, the current \code{Curated} list of examples will continue growing, so keep an eye on the GitHub repository. On that note, if you start using \solx and you have a set of data entries that you deem valuable, email me at \solxmail\ and I will consider making you a curator, so that your entries can enter the ``official'' library.

\section*{Use of AI}
The first prototype of \solx was developed by hand in 2022, for the calculations of \cite{Bobev:2022bjm}, and the core of the package is the author's: the simulation of objects with \wlref{Association}'s and the core logic of \sx{Compute}. \code{Claude Code}, running the Opus~5 and Fable~5 models, wrote the symbol reference pages of the \href{https://resources.wolframcloud.com/PacletRepository/resources/VasilDimitrov/SolutionsX/}{in-system documentation}; while the tech notes were written by the author. It also rebuilt the component engine behind \sx{Compute} towards speed-up, carried out smaller Kernel tasks and maintenance, and templated the \LaTeX{} machinery of the
present paper. The agent toolkit was developed by \code{Claude Code} after extensive discussions on its architecture. Throughout, \code{Claude Code} was treated as a collaborator: in particular, it wrote almost all of ``its
own section'': \cref{sec:agents}, reporting there how its previous instances drove \solx.

\appendix
\section{\mma notation}
\label{app:mma}

The cells of this paper use a handful of \mma idioms that a reader coming from another CAS may not have met. This appendix collects them. It does not replace the \mma documentation; it is just enough to keep reading.

\paragraph{Special characters} The indices and the products of the paper are single glyphs, each carrying a long name:
\begin{center}
\arrayrulecolor{black}
\begin{tabular}{@{}cll@{}}
\toprule
glyph & long name & role \\
\midrule
$\curly{m}, \curly{n}, \dots$ & \code{\textbackslash[ScriptM]}, \code{\textbackslash[ScriptN]}, \dots & abstract indices on a tangent bundle \\
$\mathfrak{a}, \mathfrak{b}, \dots$ & \code{\textbackslash[GothicA]}, \code{\textbackslash[GothicB]}, \dots & frame-bundle indices \\
$\mathbb{A}, \mathbb{B}, \dots$ & \code{\textbackslash[DoubleStruckCapitalA]}, \dots & spinor indices \\
$\mathfrak{I}, \mathfrak{J}, \dots$ & \code{\textbackslash[GothicCapitalI]}, \dots & indices on further inner bundles \\
$\mathcal{M}, \mathcal{N}$ & \code{\textbackslash[ScriptCapitalM]}, \dots & manifolds \\
$\otimes$ & \code{\textbackslash[CircleTimes]} & the symmetric product of forms (\cref{sec:input}) \\
$\wedge$ & \code{\textbackslash[Wedge]} & the exterior product (\xterior) \\
$\cdot$ & \code{\textbackslash[CenterDot]} & \fieldsx's product of Grassmann-odd objects \\
$\mapsto$ & \code{\textbackslash[RightTeeArrow]} & \xtensor's infix \code{IndexRule} \\
$\llcorner\!\lrcorner$ & \code{\textbackslash[UnderBracket]} & an ordinary name character, see below \\
\bottomrule
\end{tabular}
\end{center}
The role column is a convention, not a constraint: nothing in \solx ties a glyph to a role, and the assignments above are simply what the author found visually appealing. Any symbol works in any position. Typing the long name in a notebook produces the glyph the moment the name is complete; most characters also carry a short alias between two presses of \code{Esc}, listed in the documentation. The last row exists because the underscore itself is not available in names: \code{\_} is Wolfram syntax for a pattern, so the corpus notebooks use $\llcorner\!\lrcorner$ where a human would write \code{\_}. The last two rows appear in the corpus notebooks rather than in the cells of this paper.

\paragraph{Applying functions} A function of one argument can be applied in three equivalent ways: bracketed, postfix with \code{//}, and prefix with \code{@}
\begin{wlin}
Simplify[Sin[(!\free{x}!)]^2 + Cos[(!\free{x}!)]^2]
Sin[(!\free{x}!)]^2 + Cos[(!\free{x}!)]^2 // Simplify
Simplify@(Sin[(!\free{x}!)]^2 + Cos[(!\free{x}!)]^2)
\end{wlin}
\begin{wlout}
1
\end{wlout}
\begin{wlout}[1]
1
\end{wlout}
\begin{wlout}[1]
1
\end{wlout}
\noindent The paper's cells use all three, favouring \code{//} for pipelines and \code{@} for single wrappers.

\paragraph{Pure functions} \code{Simplify[\#] \&} is a function without a name: \code{\#} marks where the argument goes and the trailing \code{\&} closes the definition. It applies in the same three ways
\begin{wlin}
(Simplify[(!\wlslot{\#}!)] &)[Sin[(!\free{x}!)]^2 + Cos[(!\free{x}!)]^2]
Sin[(!\free{x}!)]^2 + Cos[(!\free{x}!)]^2 // (Simplify[(!\wlslot{\#}!)] &)
(Simplify[(!\wlslot{\#}!)] &)@(Sin[(!\free{x}!)]^2 + Cos[(!\free{x}!)]^2)
\end{wlin}
\begin{wlout}
1
\end{wlout}
\begin{wlout}[1]
1
\end{wlout}
\begin{wlout}[1]
1
\end{wlout}
\noindent and \code{/@} maps it over a list
\begin{wlin}
Simplify[(!\wlslot{\#}!)] & /@ {Sin[(!\free{x}!)]^2 + Cos[(!\free{x}!)]^2, Cosh[(!\free{x}!)]^2 - Sinh[(!\free{x}!)]^2}
\end{wlin}
\begin{wlout}[0]
{1, 1}
\end{wlout}
\noindent The slot may sit anywhere in the body, so a function of two arguments becomes a pure function of one of them by fixing the other
\begin{wlin}
D[(!\wlslot{\#}!), (!\free{x}!)] & /@ {(!\free{x}!)^2, Sin[(!\free{x}!)]}
\end{wlin}
\begin{wlout}[0]
{2 (!\free{x}!), Cos[(!\free{x}!)]}
\end{wlout}
\noindent Most of the \sxD{Apply} entries of \cref{sec:compute} are pure functions of exactly this kind.

\paragraph{A list as an argument sequence} \code{@@} feeds a list to a function as separate arguments, so the following are the same expression
\begin{wlin}
(!\free{f}!)[(!\free{a}!), (!\free{b}!), (!\free{c}!)]
(!\free{f}!) @@ {(!\free{a}!), (!\free{b}!), (!\free{c}!)}
\end{wlin}
\begin{wlout}
(!\free{f}!)[(!\free{a}!), (!\free{b}!), (!\free{c}!)]
\end{wlout}
\begin{wlout}[1]
(!\free{f}!)[(!\free{a}!), (!\free{b}!), (!\free{c}!)]
\end{wlout}
\noindent The paper meets it as \sx{RulesToChain}\code{ @@ \{...\}} in \cref{sec:compute}.

\paragraph{Previous outputs} Within a cell, \code{\%} is the last output and \code{\%\%} the one before it; a trailing \code{;} suppresses the display of a result without suppressing the result
\begin{wlin}
Sin[(!\free{x}!)]^2 + Cos[(!\free{x}!)]^2 // Simplify;
5;
\end{wlin}
\begin{wlout}[2]
6
\end{wlout}
\noindent This is what lets the verification cells of \cref{sec:verify} read as short pipelines.

\paragraph{Rules, and rules with \code{Function}} A rule \code{lhs -> rhs} rewrites, via the operator \code{/.}, exactly the pattern it names and nothing else
\begin{wlin}
(!\free{rule}!) = {(!\free{f}!)[(!\free{x}!)] -> Sin[(!\free{x}!)] + Cos[(!\free{x}!)]};
(!\free{f}!)[(!\free{x}!)] + 5 /. (!\free{rule}!)
(!\free{f}!)'[(!\free{x}!)] + 5 /. (!\free{rule}!)
\end{wlin}
\begin{wlout}[1]
5 + Cos[(!\free{x}!)] + Sin[(!\free{x}!)]
\end{wlout}
\begin{wlout}[1]
5 + (!\free{f}!)'[(!\free{x}!)]
\end{wlout}
\noindent The second line is untouched: the rule knows what \code{f[x]} becomes but not what \code{f} \emph{is}, so the derivative has nothing to act on. The cure is a rule for the head itself, with \wlref{Function} carrying the body
\begin{wlin}
(!\free{frule}!) = {(!\free{f}!) -> Function[(!\free{x}!), Sin[(!\free{x}!)] + Cos[(!\free{x}!)]]};
(!\free{f}!)[(!\free{x}!)] + 5 /. (!\free{frule}!)
(!\free{f}!)'[(!\free{x}!)] + 5 /. (!\free{frule}!)
\end{wlin}
\begin{wlout}[1]
5 + Cos[(!\free{x}!)] + Sin[(!\free{x}!)]
\end{wlout}
\begin{wlout}[1]
5 + Cos[(!\free{x}!)] - Sin[(!\free{x}!)]
\end{wlout}
\noindent Now the substitution commutes with differentiation. This is why \sx{Gen} returns rules of exactly this shape (\cref{sec:input}).

\paragraph{Associations} An association is Wolfram's dictionary, written \code{<|key -> value|>}. Values nest, and a nested value is looked up by giving the keys in sequence
\begin{wlin}
(!\free{assoc}!) = <|"a" -> 1, "b" -> <|"c" -> 2|>|>;
(!\free{assoc}!)["b", "c"]
\end{wlin}
\begin{wlout}[1]
2
\end{wlout}
\noindent Assigning into a sequence of keys creates the entry. A record \free{sol} in this paper is one association, which is why every address in its cells has the shape \code{sol[\$metric, gg, Expression]}, and why a solution is built by assigning into such addresses (\cref{sec:input}).

\section{Public symbols of \solx}
\label{app:surface}

Every public symbol of \solx, with its usage message as displayed by the
in-system documentation. Names in \textcolor{wlteal}{teal} link within
this appendix; built-ins link to the Wolfram Language reference. This
appendix is generated from the package documentation sources and is not
edited by hand.


\sxgroup{Configuration and identity}

\sxentry{D:DefaultDataDirectory}{\$DefaultDataDirectory}
\sxusage{\sxlink{D:DefaultDataDirectory}{\$DefaultDataDirectory}}{is the path used for data and notebooks when no other is set: the SolutionsX folder under \href{https://reference.wolfram.com/language/ref/\%24LocalBase.html}{\code{\$LocalBase}}.}

\sxentry{D:DataDirectory}{\$DataDirectory}
\sxusage{\sxlink{D:DataDirectory}{\$DataDirectory}}{is the path under which data and notebooks are stored and retrieved.}
\sxprose{Entries live in \sxlink{D:DataDirectory}{\code{\$DataDirectory}}/\sxlink{D:Alias}{\code{\$Alias}}/\sxarg{name}/. Set it with \sxlink{SetDataDirectory}{\code{SetDataDirectory}}.}

\sxentry{D:Alias}{\$Alias}
\sxusage{\sxlink{D:Alias}{\$Alias}}{is the sub-directory of \sxlink{D:DataDirectory}{\code{\$DataDirectory}} holding the current user's entries: their identity.}
\sxprose{\sxlink{SaveData}{\code{SaveData}} and \sxlink{DeleteData}{\code{DeleteData}} act only under it; \sxlink{GetData}{\code{GetData}} and \sxlink{OpenData}{\code{OpenData}} default to it but read any alias via "alias/name" or the Alias option. Set it once with \sxlink{SetAlias}{\code{SetAlias}}; it persists in the user configuration file.}

\sxentry{Alias}{Alias}
\sxusage{Alias}{is an option for GetData and OpenData that selects the alias to read from, equivalent to the "alias/name" form.}

\sxentry{D:Curator}{\$Curator}
\sxusage{\sxlink{D:Curator}{\$Curator}}{is True when this user curates the published Curated corpus: it unlocks working under the Curated alias and \sxlink{CurateData}{\code{CurateData}}. Set it with \sxlink{SetCurator}{\code{SetCurator}}; it persists in the user configuration file.}

\sxentry{SetAlias}{SetAlias}
\sxusage{\sxlink{SetAlias}{SetAlias}[\sxarg{"alias"}]}{sets \sxlink{D:Alias}{\code{\$Alias}}, refreshes the argument completions, and records the alias in the user configuration file so later sessions restore it.}
\sxusage{\sxlink{SetAlias}{SetAlias}[\sxarg{"alias"}, Permanent->False]}{sets it for this session only.}

\sxentry{SetDataDirectory}{SetDataDirectory}
\sxusage{\sxlink{SetDataDirectory}{SetDataDirectory}[\sxarg{"path"}]}{sets \sxlink{D:DataDirectory}{\code{\$DataDirectory}}, refreshes the argument completions, and records the path in the user configuration file so later sessions restore it.}
\sxusage{\sxlink{SetDataDirectory}{SetDataDirectory}[\sxarg{"path"}, Permanent->False]}{sets it for this session only.}

\sxentry{SetCurator}{SetCurator}
\sxusage{\sxlink{SetCurator}{SetCurator}[\wlref{True}]}{marks this user as the curator of the Curated corpus and records it in the user configuration file; \sxlink{SetCurator}{\code{SetCurator}}[\wlref{False}] removes the mark.}
\sxusage{\sxlink{SetCurator}{SetCurator}[\sxarg{flag}, Permanent->False]}{sets it for this session only.}

\sxentry{WorkAs}{WorkAs}
\sxusage{WorkAs}{[\sxarg{"owner-bot"}] switches this session to an agent identity: \$Alias is set to the given agent alias and the curator flag is dropped. The user configuration file is never written, so the switch ends with the kernel.}
\sxprose{Only aliases of the shape \sxarg{"<owner>-bot"} are accepted --- one agent alias per human, e.g. "Vasko-bot"; your own identity is set with SetAlias. A refused alias, or a switch that does not take, aborts the evaluation. Experimental, part of the agent kit.}

\sxentry{LocateData}{LocateData}
\sxusage{\sxlink{LocateData}{LocateData}[]}{points \sxlink{D:DataDirectory}{\code{\$DataDirectory}} at the data tree containing the current notebook, for this session only, then displays the package load panel and the Data status.}
\sxusage{\sxlink{LocateData}{LocateData}[\sxarg{dir}]}{does the same for an explicit entry directory path.}
\sxprose{It relocates only when the notebook sits inside a Data entry directory (\dots{}/Data/\sxarg{alias}/\sxarg{name}/) or at a SolutionsX repository root; anywhere else it leaves the configured values untouched and just displays them. It is the display line of \sxlink{NewData}{\code{NewData}}'s initialization cell.}

\sxentry{Welcome}{Welcome}
\sxusage{Welcome[]}{guides a first session: it asks where your data should live and what your alias is, records both in the user configuration, and offers the curated entries --- individually selectable --- to copy under your alias.}
\sxprose{Without a front end it prints the same steps as commands. Welcome[] never touches the curator flag and can be re-run at any time.}

\sxentry{SolutionsXHelp}{SolutionsXHelp}
\sxusage{SolutionsXHelp[]}{opens the SolutionsX guide page in the Documentation Center.}

\sxgroup{Objects and their keys}

\sxentry{Gen}{Gen}
\sxusage{\sxlink{Gen}{Gen}[\sxarg{sol}[\$function,\sxarg{key}]]}{gives the rule sending the scalar function \sxarg{key} to its stored Expression, for use with /.}
\sxusage{\sxlink{Gen}{Gen}[\sxarg{sol}[\$form,\sxarg{key}]]}{gives the rule sending the form's Symbol to its stored Expression.}
\sxusage{\sxlink{Gen}{Gen}[\sxarg{sol}[\$rule,\sxarg{key}]]}{gives the stored Expression of the rule entry \sxarg{key}.}
\sxprose{\sxlink{Gen}{\code{Gen}} gives \{\} when the entry is absent, so it is safe to apply unconditionally.}

\sxentry{Dimension}{Dimension}
\sxusage{Dimension}{is a key of \$manifold and \$bundle objects: the dimension, a positive integer.}

\sxentry{Index}{Index}
\sxusage{Index}{is a key of \$manifold, \$frame, \$spinStructure and \$bundle objects: the abstract indices to register, as a list of symbols.}

\sxentry{Name}{Name}
\sxusage{Name}{is a key of \$manifold and \$chart objects: a string from which SaveData composes the entry name.}

\sxentry{D:auto}{\$auto}
\sxusage{\$auto}{is a key of most objects: slots for the symbols xAct creates alongside the definition, each carrying a Routine (the Chain recipe Compute follows) and its computed Value.}

\sxentry{ValidateObject}{ValidateObject}
\sxusage{\sxlink{ValidateObject}{ValidateObject}[\sxarg{object},\sxarg{expr},\sxarg{types}]}{checks the association \sxarg{expr} against the template of \sxarg{object}, and returns it unchanged if it passes.}
\sxusage{\sxlink{ValidateObject}{ValidateObject}[\$solution,\sxarg{sol},\sxarg{types}]}{checks a whole solution, object type by object type.}
\sxprose{\sxarg{types} is a list of key selectors, normally \{PropKeysOf, OptKeysOf\}. Failure issues a message and throws \sxarg{object}.}

\sxentry{IncludeTo}{IncludeTo}
\sxusage{\sxlink{IncludeTo}{IncludeTo}[\sxarg{sol}[\sxarg{object}],\sxarg{assoc}]}{validates \sxarg{assoc}, stores it under \sxarg{object}, and makes the corresponding xAct definition.}
\sxusage{\sxlink{IncludeTo}{IncludeTo}[\sxarg{sol}[\sxarg{object}],\{\sxarg{assoc}..\}]}{does the same for several entries in order.}
\sxprose{The keys of \sxarg{assoc} are those of \wlref{Options}[\sxarg{object}]; missing keys take their default. Any entry already under the same key is unloaded first.}

\sxentry{DropFrom}{DropFrom}
\sxusage{\sxlink{DropFrom}{DropFrom}[\sxarg{sol}[\sxarg{object}],\sxarg{key}]}{undefines the entry and removes it from \sxarg{sol}; a list of keys is also accepted.}
\sxprose{Unlike \sxlink{Unload}{\code{Unload}}, \sxlink{DropFrom}{\code{DropFrom}} forgets the stored association as well.}

\sxentry{MergeWith}{MergeWith}
\sxusage{\sxlink{MergeWith}{MergeWith}[\sxarg{sol},\sxarg{assoc}]}{includes and loads every entry of the solution association \sxarg{assoc} into \sxarg{sol}.}
\sxprose{It refuses, with a message, if the two share any symbol or any key.}

\sxentry{Instantiate}{Instantiate}
\sxusage{\sxlink{Instantiate}{Instantiate}[\sxarg{sol}]}{clears \sxarg{sol} and makes it an empty solution: an association over \$Objects, carrying the UpValues of \sxlink{Load}{\code{Load}}, \sxlink{Unload}{\code{Unload}}, \sxlink{IncludeTo}{\code{IncludeTo}}, \sxlink{DropFrom}{\code{DropFrom}} and \sxlink{MergeWith}{\code{MergeWith}}.}
\sxusage{\sxlink{Instantiate}{Instantiate}[\sxarg{sol},\sxarg{assoc}]}{does the same and then merges \sxarg{assoc} into it.}

\sxgroup{Building expressions}

\sxentry{MakeArray}{MakeArray}
\sxusage{\sxlink{MakeArray}{MakeArray}[\sxarg{expr}]}{gives the component array of \sxarg{expr}, which may be a line element written with \wlref{CircleTimes} or a differential form written with \wlref{Wedge}.}
\sxprose{The chart is deduced from the Diff's appearing in \sxarg{expr}. It is the inverse of \sxlink{MakeMetric}{\code{MakeMetric}} and \sxlink{MakeForm}{\code{MakeForm}}.}

\sxentry{MakeMetric}{MakeMetric}
\sxusage{\sxlink{MakeMetric}{MakeMetric}[\sxarg{chart}][\sxarg{array}]}{gives the line element of the symmetric matrix \sxarg{array} in \sxarg{chart}, as a sum of \wlref{CircleTimes} of coordinate differentials.}

\sxentry{MakeForm}{MakeForm}
\sxusage{\sxlink{MakeForm}{MakeForm}[\sxarg{chart}][\sxarg{array}]}{gives the differential form of the totally antisymmetric \sxarg{array} in \sxarg{chart}, as a sum of \wlref{Wedge} of coordinate differentials.}

\sxentry{RulesToChain}{RulesToChain}
\sxusage{\sxlink{RulesToChain}{RulesToChain}[\sxarg{a} -> \sxarg{b} -> \sxarg{c}]}{gives \{\sxarg{a} -> \sxarg{b}, \sxarg{b} -> \sxarg{c}\}, the form \sxlink{Compute}{\code{Compute}} expects for a raising or lowering ladder.}

\sxentry{ExpandForms}{ExpandForms}
\sxusage{\sxlink{ExpandForms}{ExpandForms}[\sxarg{sol}][\sxarg{expr}]}{rewrites every differential form of \sxarg{sol} appearing in \sxarg{expr} as its companion tensor contracted with a wedge of coordinate differentials, and expands the exterior derivative of every degree-zero expression in the same sweep.}

\sxentry{WedgeCoeff}{WedgeCoeff}
\sxusage{\sxlink{WedgeCoeff}{WedgeCoeff}[\sxarg{sol}][\sxarg{expr}]}{extracts the antisymmetric component tensor of a form written in wedges of coordinate differentials, as the lowest-rank equivalent: a form of degree p above half the dimension d is Hodge-dualized and expanded first, and the extraction runs on the dual, giving the component tensor of the Hodge dual with d-p free indices.}
\sxprose{For p at most d/2 the result is the plain component tensor, so \sxlink{WedgeCoeff}{\code{WedgeCoeff}} inverts \sxlink{ExpandForms}{\code{ExpandForms}}: the companion tensor comes back exactly.}
\sxprose{Reconstructing a dualized form from its result costs the double-dual sign (-1)\^{}(p(d-p)) sign(det g) --- in Lorentzian signature the volume form extracts to minus its coefficient. That sign is mathematics, not a convention.}
\sxprose{The result is presented in the order of the manifold's Index list: the frees introduced by the extraction take the first names not already free in \sxarg{expr}, and ScreenDollarIndices renames the dummies to the following unused ones. If the registered list is exhausted mid-extraction, new indices are generated and registered, so the extraction always succeeds.}
\sxprose{Only ContractMetric is applied to the result; simplification is left to the caller.}

\sxentry{ComputeDiffs}{ComputeDiffs}
\sxusage{\sxlink{ComputeDiffs}{ComputeDiffs}[\sxarg{chart}]}{installs the rules that canonicalise, and annihilate repeated, wedge products of the coordinate differentials of \sxarg{chart}.}

\sxentry{BasisOfVBundle}{BasisOfVBundle}
\sxusage{\sxlink{BasisOfVBundle}{BasisOfVBundle}[\sxarg{bundle}]}{gives the first basis or chart defined over \sxarg{bundle}, or \wlref{None} if there is none.}
\sxprose{It issues a message and throws \href{https://reference.wolfram.com/language/ref/\%24Failed.html}{\code{\$Failed}} if \sxarg{bundle} is not a vector bundle.}

\sxentry{ToBases}{ToBases}
\sxusage{\sxlink{ToBases}{ToBases}[\sxarg{expr}]}{rewrites every abstract index of \sxarg{expr} in the basis belonging to its vector bundle.}
\sxprose{The option Not -> \{\sxarg{basis}..\} leaves those bases untouched.}

\sxentry{ToArray}{ToArray}
\sxusage{\sxlink{ToArray}{ToArray}[\sxarg{expr}]}{gives the explicit component array of \sxarg{expr}, by taking it to bases, tracing dummies and resolving component values.}
\sxprose{The option Not -> \{\sxarg{basis}..\} leaves those bases untouched.}

\sxgroup{Computing}

\sxentry{Compute}{Compute}
\sxusage{\sxlink{Compute}{Compute}[\sxarg{sol}[\sxarg{object},\sxarg{key},\$auto,\sxarg{autoKey}]]}{computes the components of \sxarg{autoKey} by walking its stored Chain, and stores them under Value.}
\sxusage{\sxlink{Compute}{Compute}[\sxarg{sol}[\sxarg{object},\sxarg{key}]]}{does this for every automatic tensor of the entry.}
\sxprose{A Chain entry \wlref{HoldForm}[\sxarg{expr}] -> \{\sxarg{slots}\} takes the components from \sxarg{expr}; an entry \{\sxarg{slots}\} -> \{\sxarg{slots}\} raises or lowers indices with the metric. Nothing is computed if the Chain is unset.}
\sxprose{Options are Chain, which overrides and stores the recipe, Using, which chooses the metric of each bundle, and Apply, which defaults to \sxlink{D:Apply}{\code{\$Apply}}.}
\sxprose{Compute checks the Chain's slot specification against the object, reports an Apply key that is neither \wlref{Map} nor \wlref{ParallelMap}, and refuses to raise a slot whose vector bundle has no metric.}

\sxentry{D:Apply}{\$Apply}
\sxusage{\sxlink{D:Apply}{\$Apply}}{is the default value of the Apply option of \sxlink{Compute}{\code{Compute}}.}
\sxprose{It is a list of rules whose keys are \wlref{Map} or \wlref{ParallelMap} and whose values are applied to computed components; the default is \{\wlref{Map} -> \wlref{Together}, \wlref{ParallelMap} -> \wlref{Simplify}\}.}

\sxgroup{Loading and unloading}

\sxentry{Load}{Load}
\sxusage{\sxlink{Load}{Load}[\sxarg{sol}]}{installs the UpValues of \sxarg{sol}, then makes the xAct definitions of every entry it holds.}
\sxusage{\sxlink{Load}{Load}[\sxarg{sol}[\sxarg{object}],\sxarg{key}]}{loads a single entry; a list of keys is also accepted.}
\sxprose{\sxlink{Load}{\code{Load}} is the step that turns an association returned by \sxlink{GetData}{\code{GetData}} into live xAct definitions.}

\sxentry{Unload}{Unload}
\sxusage{\sxlink{Unload}{Unload}[\sxarg{sol}]}{undefines every xAct object of \sxarg{sol}, in reverse order of construction, keeping the stored association.}
\sxusage{\sxlink{Unload}{Unload}[\sxarg{sol}[\sxarg{object}],\sxarg{key}]}{undefines a single entry; a list of keys is also accepted.}
\sxprose{\sxlink{Unload}{\code{Unload}} does not delete the entry: use \sxlink{DropFrom}{\code{DropFrom}} for that.}

\sxentry{Resurrect}{Resurrect}
\sxusage{\sxlink{Resurrect}{Resurrect}[\sxarg{symbol}]}{rebinds any symbol removed by an xAct Undef, by round-tripping \sxarg{symbol} through \wlref{Compress} and replacing Removed[\sxarg{name}] by \wlref{Symbol}[\sxarg{name}].}

\sxentry{Revive}{Revive}
\sxusage{\sxlink{Revive}{Revive}[\sxarg{sol}]}{applies \sxlink{Resurrect}{\code{Resurrect}} throughout \sxarg{sol}, to every symbol in the Global` context and to \href{https://reference.wolfram.com/language/ref/\%24Assumptions.html}{\code{\$Assumptions}}. It is called after each unload.}

\sxentry{ResumeAs}{ResumeAs}
\sxusage{\sxlink{ResumeAs}{ResumeAs}[\sxarg{sol}]}{restores the current notebook's entry into \sxarg{sol} (\sxlink{GetData}{\code{GetData}}[] followed by \sxlink{Load}{\code{Load}}) when nothing has been loaded in this kernel yet, and rebuilds the Load banner of everything loaded so far otherwise.}
\sxprose{Put it at the top of a notebook section: after a kernel quit it resumes from disk; in a full top-to-bottom run it only re-shows the banner, so the same notebook supports both workflows.}

\sxgroup{Storing and retrieving}

\sxentry{NewData}{NewData}
\sxusage{\sxlink{NewData}{NewData}[]}{creates a notebook holding the initialization cell that loads SolutionsX.}
\sxprose{The cell carries no user identity and no absolute paths: \sxlink{LocateData}{\code{LocateData}}[] finds the data root relative to the notebook's own position; \sxlink{D:Alias}{\code{\$Alias}} always comes from the user configuration.}

\sxentry{SaveData}{SaveData}
\sxusage{\sxlink{SaveData}{SaveData}[\sxarg{sol}]}{writes \sxarg{sol} and the evaluation notebook into \sxlink{D:DataDirectory}{\code{\$DataDirectory}}/\sxlink{D:Alias}{\code{\$Alias}}/\sxarg{name}/, and echoes the destination.}
\sxprose{The name is built from the first manifold's Name and Dimension, the first metric's Signature and the first chart's Name: an entry with a chart is a Sol, one without is a Thr.}
\sxprose{Every save stamps \sxarg{sol}[\$info]: SolutionsX version, the saving alias (LastEdit), date, and the Wolfram and xAct versions. The stamp is machine-written and invisible to validation.}

\sxentry{Overwrite}{Overwrite}
\sxusage{Overwrite}{is an option for CopyData and CurateData: Automatic (default) asks per existing entry in a front end and skips headless; True replaces silently; False always skips.}

\sxentry{GetData}{GetData}
\sxusage{\sxlink{GetData}{GetData}[\sxarg{name}]}{gives the stored solution association of the entry \sxarg{name} under the user's own alias, validated but not loaded.}
\sxusage{\sxlink{GetData}{GetData}[\sxarg{"alias/name"}]}{(or the option Alias->\sxarg{"alias"}) reads the entry of another alias.}
\sxusage{\sxlink{GetData}{GetData}[]}{takes the alias and name from the directory containing the current notebook.}
\sxprose{Follow \sxlink{GetData}{\code{GetData}} with \sxlink{Load}{\code{Load}} to make the definitions.}

\sxentry{OpenData}{OpenData}
\sxusage{\sxlink{OpenData}{OpenData}[\sxarg{name}]}{opens the notebook of the entry \sxarg{name} under the user's own alias.}
\sxusage{\sxlink{OpenData}{OpenData}[\sxarg{"alias/name"}]}{(or the option Alias->\sxarg{"alias"}) opens another alias's entry.}

\sxentry{ShowData}{ShowData}
\sxusage{\sxlink{ShowData}{ShowData}[]}{displays an interactive browser of the data tree: one button per alias --- your own first, marked with a pencil --- and the entries of the chosen alias with their \sxlink{OpenData}{\code{OpenData}}, CopyName, \sxlink{CopyData}{\code{CopyData}} and \sxlink{DeleteData}{\code{DeleteData}} actions.}
\sxprose{Every entry shows the name \sxlink{GetData}{\code{GetData}} takes, and CopyName copies it as a quoted string, ready to paste. The search field filters across every alias as you type. Without a front end ShowData[] prints the same tree as text.}

\sxentry{CopyData}{CopyData}
\sxusage{\sxlink{CopyData}{CopyData}[\sxarg{"alias/name"}]}{copies the entry \sxarg{name} of another alias into your own alias, notebook and all; the source is never touched.}
\sxusage{\sxlink{CopyData}{CopyData}[\sxarg{"alias"}]}{copies every entry of that alias into yours, one by one.}

\sxentry{CurateData}{CurateData}
\sxusage{\sxlink{CurateData}{CurateData}[\sxarg{"name"}]}{publishes the entry \sxarg{name} of your own alias into the Curated corpus. Curator-only; enable with \sxlink{SetCurator}{\code{SetCurator}}.}

\sxentry{DeleteData}{DeleteData}
\sxusage{\sxlink{DeleteData}{DeleteData}[\sxarg{name}]}{permanently deletes the entry directory \sxarg{name} and all of its contents.}
\sxprose{\sxlink{DeleteData}{\code{DeleteData}} acts only under the user's own alias; entries of other aliases cannot be deleted.}

\section{Configuring the agent toolkit}
\label{app:kit}

The toolkit of \cref{sec:kit} is configured by one flat file, \code{agent/kit.conf}: \code{KEY=VALUE} lines, integers only. The file is parsed, never sourced, so a malformed line is ignored (and warned about) rather than executed. Four layers override each other, highest first: \code{sxk set KEY value}, which lives for one session and is surfaced to the agent like a user note until acknowledged; the environment variable \code{SXK\_KEY}, for scripted runs; the file itself, holding the owner's standing preferences; and the built-in defaults. \code{sxk config} prints every effective value together with the layer it came from.

Each lever is one of two kinds. An \emph{enforced} lever is acted on by \code{sxk} itself: changing it changes behaviour. An \emph{advisory} lever is only displayed by \code{sxk}; \code{LADDER.md} instructs the agent to act on it, and nothing can force compliance, by design.

\begin{table}[h]
\centering
\small
\begin{tabular}{@{}llrl@{}}
\toprule
lever & kind & default & governs \\
\midrule
\code{TRY\_SECONDS}   & enforced & 120  & the time budget of \code{sxk try} when none is given \\
\code{DO\_TIMEOUT}    & enforced & 600  & how long \code{sxk do} blocks before returning control \\
\code{WATCH\_INTERVAL}& enforced & 5    & the polling interval of \code{sxk watch} \\
\code{WATCH\_MAX}     & enforced & 600  & total watch time before \code{sxk watch} gives up waiting \\
\code{WAIT\_MAX}      & enforced & 3600 & the timeout of \code{sxk wait} \\
\code{STALL\_SHAPES}  & advisory & 3    & failed shapes on one obstacle before the agent must stop \\
\code{STALL\_MINUTES} & advisory & 30   & minutes without a new verified fact before the same \\
\bottomrule
\end{tabular}
\caption{The seven levers of \code{agent/kit.conf}. Defaults in seconds
where the name says so. Three behaviours are worth spelling out. A
\code{try} that expires cuts off the expression but the kernel, with all
its state, survives: a wrong-shaped attempt costs \code{TRY\_SECONDS},
not a silent half-hour. A \code{do} that times out returns the prompt to
the agent while the submission keeps running, and a \code{watch} that
ends is a decision point, not an abort. \code{STALL\_MINUTES} is also
the one advisory lever with an enforced display: a busy kernel whose
compute pulse is older than it is flagged as stale by \code{sxk status}
and \code{sxk watch}, as a report, not a verdict.}
\label{tab:kitconf}
\end{table}

\newpage
\bibliographystyle{JHEP}
\bibliography{refs}

\end{document}